\documentclass[
  reprint,
  aps,
  prx,
  amsmath,
  amssymb,
  floatfix,
  superscriptaddress,
  longbibliography
]{revtex4-2}

\usepackage[utf8]{inputenc}
\usepackage[T1]{fontenc}
\usepackage{microtype}
\usepackage{graphicx}
\usepackage[table]{xcolor}
\usepackage{placeins}
\usepackage{needspace}
\usepackage{physics}
\usepackage[colorlinks,linkcolor=blue,citecolor=blue,urlcolor=blue]{hyperref}

\usepackage{tikz}

\newcommand{\mean}[1]{\ensuremath{\langle #1\rangle}}

\begin{document}

\title{Avalanches and Aging Driven by Parametric Amplification in Kerr Oscillator Networks}

\author{Alejandro S. Gómez}
\email{alejandros.gomez@uam.es}
\affiliation{Departamento de Física Teórica de la Materia Condensada and Condensed Matter Physics Center (IFIMAC),
Universidad Autónoma de Madrid, E-28049 Madrid, Spain}
\author{Javier del Pino}
\email{j.delpino@uam.es}
\affiliation{Departamento de Física Teórica de la Materia Condensada and Condensed Matter Physics Center (IFIMAC),
Universidad Autónoma de Madrid, E-28049 Madrid, Spain}

\begin{abstract}
Gain, loss, and nonlinearity shape the phase-space landscape of driven-dissipative systems, such as Kerr parametric oscillator (KPO) networks realized in superconducting, nanomechanical, and photonic arrays.
Networks of KPOs exploit the stationary states of this landscape for Ising machines, bosonic error-correcting codes, and sensing.
Yet how transients reach these states from a given initial condition remains largely unexplored.
Here we show that KPO networks with nonlinear couplings can
develop avalanches in their deterministic transient dynamics: rapid cascades of amplification between modes reminiscent of the relaxation avalanches of glassy systems.
The mechanism is that parametric gain first activates the modes whose detuning it can overcome, and their nonlinear frequency shifts then pull dormant modes into the gain window one at a time.
Dormant modes keep decaying while they wait for their turn, so every activation starts from a smaller seed and the intervals between activations grow geometrically. The network thus ages: the longer it has evolved, the slower it changes, and activations accumulate only logarithmically in time. In a 20-mode network, each dwell lasts about twice as long as the previous one.
\end{abstract}

\maketitle

\section{Introduction}

Synthetic driven-dissipative nonlinear systems exhibit synchronization~\cite{acebron2005kuramoto}, multistability~\cite{feudel1997multistability}, and pattern formation~\cite{cross1993pattern}, all controlled by tunable external parameters.
In nanomechanics, photonics, superconducting circuits, and polaritons, drive, hopping, and Kerr and cross-Kerr interactions all act as tunable knobs~\cite{slim2026strongnanomechanicalduffing,marandi2014network,kounalakis2018tuneablehopping,ye2025nearultrastrong,kim2026tunablemodelesscrosskerr,amo2016excitonpolaritons,peropadre2013tunablecoupling,alvarez2024biasedising,vonlupke2024engineering}.
In networks, coupling multiplies the coexisting collective states~\cite{tanaka2011multistable,sosa2021multistability,wiley2006sync}, and transients can linger near unstable states and switch between them along rapid connections, so that the route to the final state passes through long-lived intermediate states~\cite{hansel1993clustering,kori2001slow,timme2002prevalence,ashwin2007dynamics}.
The same competition among long-lived states underlies slow relaxation in glasses~\cite{bouchaud1998outofequilibrium,berthier2011theoretical} and frustrated elastic networks~\cite{shohat2025emergent}, memory in assemblies of bistable hysterons~\cite{keim2019memory,shohat2025geometric,paulsen2026mechanical,baconnier2025dynamic}, and sequential switching in ecological communities and neural circuits~\cite{reichenbach2007mobility,szolnoki2014cyclic,swailem2025parametric,rabinovich2006dynamical,rabinovich2011robust}.

A parametric drive creates excitations in pairs and squeezes the field, amplifying one quadrature and suppressing the fluctuations of its conjugate~\cite{clerk2010introduction}, a phase-sensitive gain that opens a direct route to non-Hermitian physics.
In linear dynamics, squeezing reshapes spectra and band structures~\cite{peano2016topological}, from level attraction~\cite{PhysRevA.98.023841} to point-gap topology~\cite{wan2023quantumsqueezing}, chiral transport, and directional amplification~\cite{peano2016topological,mcdonald2018phase,slim2024bosonickitaev,wanjura2020topological,wanjura2021correspondence,gomezleon2023drivendissipative}.
Nonlinearity changes the problem qualitatively: above threshold, linear gain would make the amplitude grow without bound, whereas in Kerr parametric oscillators (KPOs) the Kerr shift saturates this growth and stabilizes two period-doubled states of opposite phase.
These phase states encode Ising spins, so coupled KPOs operate as coherent Ising machines~\cite{wang2013coherent,marandi2014network,goto2016bifurcation,yamamoto2017coherent,kanao2021highaccuracy,alvarez2024biasedising} and form the protected manifold of cat-code bosonic quantum error correction~\cite{puri2019stabilized,grimm2020stabilization,kwon2022autonomous}.
Mixed phase states, finite-amplitude branches, and oscillator instabilities make the stationary landscape of coupled KPOs far richer than the $2^N$ spin configurations~\cite{ameye2025parametric,heugel2021ising,heugel2024proliferation,calvanesestrinati2019theory,bello2019persistent}.
Although their stationary fixed-point (FP) phases are known~\cite{ameye2025parametric}, whether, how fast, and along which transient pathways these networks reach them from prescribed initial states remains largely unexplored.

Avalanches are common in slowly driven nonlinear systems, where stress, activity, or imbalance accumulates until a threshold is reached and is then released suddenly~\cite{bak1987selforganized,sethna2001crackling}.
In the canonical sandpile, gradual loading brings a site to threshold, and its sudden discharge transfers load to its neighbors, which can discharge in turn and produce a cascade~\cite{bak1987selforganized}.
Similar intermittent dynamics, with bursts of activity separated by long intervals of slow evolution, appears in elastic networks tuned close to mechanical instability~\cite{shohat2025emergent}, in crumpled sheets~\cite{matan2002crumpling,lahini2023crackling,shohat2023logarithmic,korchinski2025thermal}, and in metamaterials built from hysterons, where the switching sequence is fixed by design~\cite{shohat2025geometric,paulsen2026mechanical,baconnier2025dynamic}.
Related cascades occur in quantum many-body systems, from avalanches of resonant tunneling in a metastable, cavity-coupled Mott insulator~\cite{hruby2018metastability} to the spreading of a thermal inclusion through a many-body localized system~\cite{leonard2023probing}.
Each avalanche can delay the following ones, so that the dynamics slows down with the time elapsed since preparation. This history-dependent slowing is called \emph{aging}~\cite{bouchaud1998outofequilibrium,berthier2011theoretical}.
In these examples, avalanches release energy accumulated by loading. Active matter instead injects energy continuously at the scale of its constituents~\cite{berthier2013nonequilibrium,janssen2019active}, and dense active assemblies age~\cite{mandal2020multiple,janzen2022aging} and relax through avalanches~\cite{keta2023intermittent,morse2021direct}.
What becomes of avalanches and aging when dissipation is intertwined with amplification?

\begin{figure}[t]
 \centering
 \includegraphics[width=0.48\textwidth]{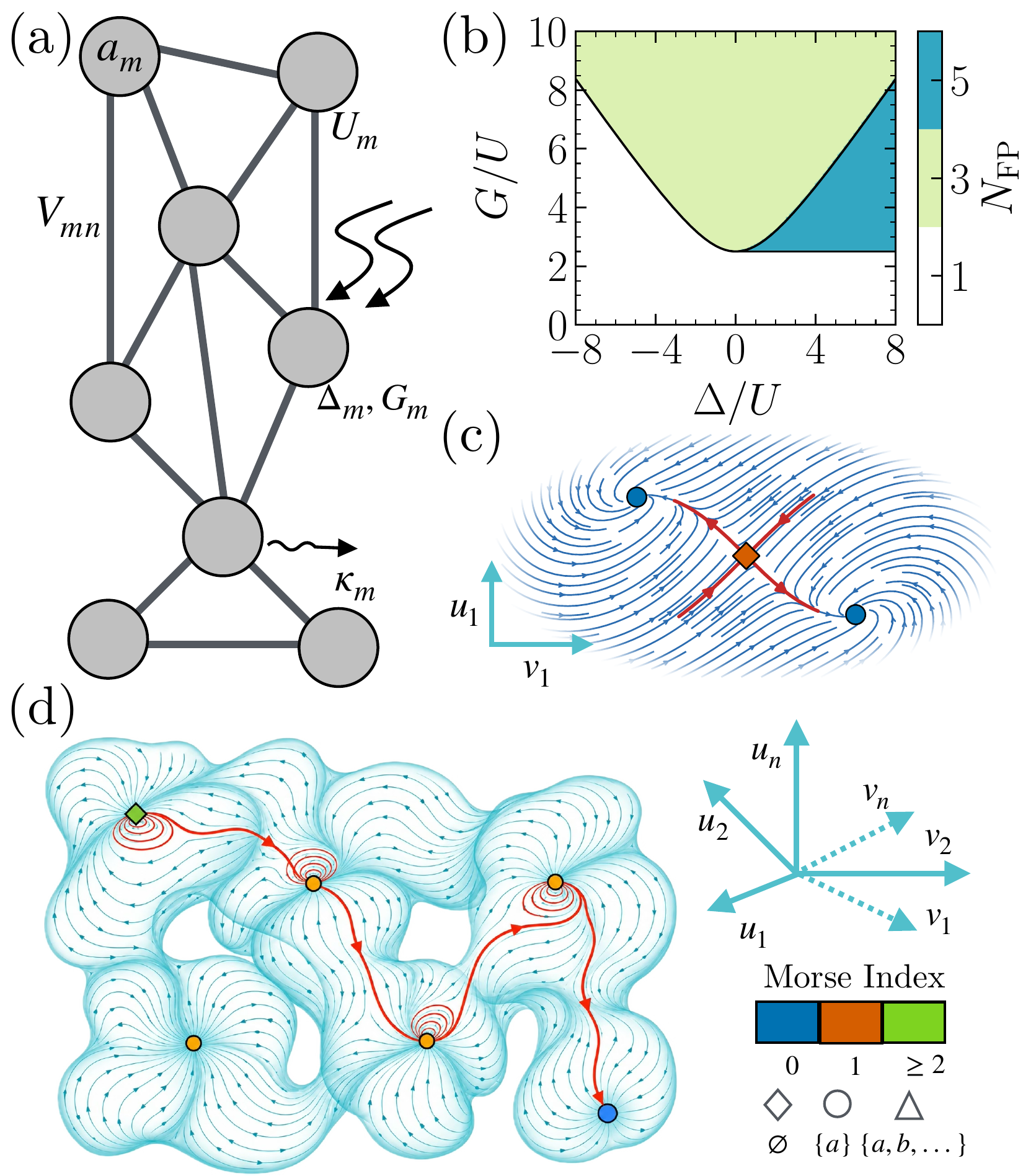}
  \caption{
    (a)~Network of coupled driven-dissipative nonlinear resonators with detunings $\Delta_m$, two-photon drives $G_m$, Kerr nonlinearities $U_m$, loss rates $\kappa_m$, and cross-Kerr couplings $V_{m,n}$.
    (b)~Mean-field phase diagram of a single KPO in the $(\Delta/U,G/U)$ plane at $\kappa/U=5$, with one, three, or five FPs. Here $U=U_m$.
    (c)~Single-KPO flow in the three-FP regime at $(\Delta/U,G/U)=(0,3)$, where the unstable branches of the saddle at the origin lead to the two stable phase states.
    (d)~Schematic of saddles in a higher-dimensional network, whose attracting directions guide the approach to modal plateaus and whose repelling ones set the escape.
    }
    \label{fig:1}
\end{figure}

We show that parametrically driven Kerr networks realize \textit{modal avalanches}: cascades of activity between modes that develop as parametric amplification drives the network out of equilibrium toward a nonequilibrium steady state, rather than as it relaxes toward equilibrium.
A sudden increase of the parametric drive from near the empty state amplifies only the modes whose detuning lies within the gain window, where gain exceeds loss. These modes ring up until the Kerr shift saturates them and become active, i.e., populated, while the others remain dormant; gain-dissipative simulators exploit the same selection, populating first the configuration with the lowest threshold~\cite{berloff2017realizing,kalinin2018global}.
An avalanche then proceeds in three steps: (i)~the active modes shift the resonance frequency of a dormant mode through the cross-Kerr interaction; (ii)~once this mode enters its gain window, its small residual amplitude, the \emph{seed}, grows; (iii)~the newly active mode shifts the remaining dormant modes and can trigger further activations.
Between activations, the active modes hold nearly constant intensities on a modal plateau, and the network stays close to a saddle until it is repelled toward a dormant mode [Fig.~\ref{fig:1}]; we call the duration of this stay the dwell time.

These avalanches age: the longer the network has evolved, the longer it waits before the next activation, because dormant seeds decay during each plateau and later activations start from smaller seeds. In the $N=20$ avalanche studied here, activations accumulate approximately logarithmically in time, without disorder.
The inherited seeds also retain a quantitative memory of the initial amplitude: the decay and amplification rates at the saddles predict the leading dependence of each dwell on the initial amplitude, linking the fixed-point spectra to the preparation dependence of the dwell times.
We also find cyclic avalanches, in which the network returns to the same plateaus in a closed loop~\cite{may1975nonlinear,guckenheimer1988structurally,krupa1995asymptotic}; they age as well, each loop lasting longer than the last.

Section~\ref{sec:model} introduces the model, Sec.~\ref{sec:chains} establishes the mechanism in a two-mode network, and Sec.~\ref{sec:landscapes} characterizes aging in many-mode avalanches and predicts its dependence on the initial perturbation from the fixed-point spectra. Section~\ref{sec:discussion} discusses the scope of this deterministic aging mechanism and concludes with an outlook.

\section{KPO network model}\label{sec:model}

\paragraph*{Network model.}
We consider a network of $N$ parametrically driven KPOs $\hat a_m$, which we also call modes, with $m=1,\ldots,N$, detunings $\Delta_m$, onsite nonlinearities $U_m$, two-photon drives $G_m$, cross-Kerr couplings $V_{m,n}$, and single-photon damping rates $\kappa_m$ [Fig.~\ref{fig:1}(a)].
Each mode is driven parametrically near its own two-photon resonance, $\omega_{p,m}\simeq2\omega_{0,m}$. In the frame rotating for each mode at half its drive frequency, $\omega_{p,m}/2$, the density matrix $\hat\rho$ obeys the Lindblad master equation
\begin{equation}
  \partial_t\hat\rho = -i[\hat H,\hat\rho] + \sum_{m = 1}^{N} \kappa_m \mathcal D[\hat a_m]\hat\rho,
  \label{eq:master_equation_general}
\end{equation}
where $\mathcal D[\square]\hat\rho=\square\hat\rho\square^\dagger-\frac12\{\square^\dagger\square,\hat\rho\}$ is the dissipator, and the Hamiltonian is
\begin{align}
  \hat H =& \sum_{m = 1}^{N} \left[ -\Delta_m \hat n_m +\frac{U_m}{2}\hat a_m^{\dagger2}\hat a_m^2 +\frac{G_m}{2}(\hat a_m^{\dagger2} + \hat a_m^2) \right] \nonumber\\
  & + \frac12\sum_{m = 1}^{N}\sum_{n\in\partial m} V_{m,n}\hat n_m\hat n_n,
  \label{eq:general_hamiltonian}
\end{align}
where the two-photon drive arises from a periodic modulation of the bare resonance frequency $\omega_{0,m}$ of mode $m$ at $\omega_{p,m}$, so that $\Delta_m=\omega_{p,m}/2-\omega_{0,m}$. $\hat n_m=\hat a_m^\dagger \hat a_m$ denotes the number operator, and $\partial m$ is the set of modes coupled to mode $m$.
Equation~\eqref{eq:general_hamiltonian} retains the terms that are static in these rotating frames and describes weakly nonlinear, parametrically driven modes in nanomechanical, optical, and superconducting platforms~\cite{eichler2023classical,grimm2020stabilization}. We use positive Kerr coefficients; Appendix~\ref{app:model_validity} gives a Josephson-circuit realization and the rotating-wave conditions under which this model holds.

\paragraph*{FPs and stability.}
We focus on the high-occupation limit, in which the Kerr shift per excitation is small and the active modes hold many excitations (Appendix~\ref{app:model_validity}). Correlations can then be neglected, $\mean{\hat n_n\hat a_m}\simeq|\alpha_n|^2\alpha_m$ with $\alpha_m=\mean{\hat a_m}$, and Eq.~\eqref{eq:master_equation_general} gives the flow
\begin{equation}
    i\dot\alpha_m=\Bigl(-\Delta_m-i\frac{\kappa_m}{2}+\sum_{n}V_{m,n}|\alpha_n|^2\Bigr)\alpha_m+G_m\alpha_m^\ast ,
    \label{eq:mf_general_complex}
\end{equation}
where $V_{m,m}=U_m$ and $V_{m,n}=V_{n,m}$, with $V_{m,n}=0$ for uncoupled modes. The amplitudes $\boldsymbol\alpha=(\alpha_1,\ldots,\alpha_N)^T$, with $\alpha_m=u_m+iv_m$, define a $2N$-dimensional real phase space. The flow has a $(\mathbb Z_2)^N$ symmetry under independent sign flips $\alpha_m\to-\alpha_m$.
An FP $\boldsymbol\alpha^{(j)}$ is a stationary solution of Eq.~\eqref{eq:mf_general_complex}. We define $\delta\hat a_m=\hat a_m-\alpha_m^{(j)}$. Linearizing around it, the fluctuation Nambu vector $\delta\hat{\boldsymbol\xi}=(\delta\hat a_1,\ldots,\delta\hat a_N,\delta\hat a_1^\dagger,\ldots,\delta\hat a_N^\dagger)^T$ obeys
\begin{equation}
    \partial_t\,\delta\hat{\boldsymbol\xi} = \mathcal J^{(j)}\,\delta\hat{\boldsymbol\xi},
    \label{eq:jacobian_general}
\end{equation}
with the Jacobian $\mathcal J^{(j)}$ of the flow at FP $j$.
We call the number of unstable directions of an FP, namely the number of eigenvalues of $\mathcal J^{(j)}$ with positive real part counted with algebraic multiplicity, its Morse index $\mu^{(j)}$~\cite{milnor1963morse}. An FP is hyperbolic when no eigenvalue has zero real part; then $\mu^{(j)}=0$ denotes a stable sink and $0<\mu^{(j)}<2N$ a saddle. For uniform loss, $\kappa_m=\kappa$, the spectrum gives $\mu^{(j)}\le N$ (Appendix~\ref{app:supports}). Note that the vacuum $\varnothing$, $\boldsymbol\alpha=\boldsymbol 0$, is an FP for all parameters.

\paragraph*{Single-mode parametric instability.}
For a single mode $m$ whose neighbors remain in the vacuum, the block of the Jacobian at the vacuum, acting on $(\delta\hat a_m,\delta\hat a_m^\dagger)^T$, is
\begin{equation}
\mathcal J_m^{(\varnothing)}=
\begin{pmatrix}
i\Delta_m-\frac{\kappa_m}{2} & -iG_m \\
iG_m & -i\Delta_m-\frac{\kappa_m}{2}
\end{pmatrix},
\label{eq:single_kpo_origin_jacobian}
\end{equation}
with eigenvalues $\lambda_\pm^m(\varnothing)=-\kappa_m/2\pm\sqrt{G_m^2-\Delta_m^2}$, where the superscript names the perturbed mode and the argument the FP.
The vacuum loses stability when the parametric gain overcomes loss and detuning, $|\Delta_m|<\chi_m$, where
\begin{equation}
    \chi_m=\sqrt{G_m^2-\kappa_m^2/4}
    \label{eq:single_kpo_threshold}
\end{equation}
is real for $G_m\geq\kappa_m/2$ and measures the gain in excess of the loss; we call $|\Delta_m|<\chi_m$ the \emph{gain window} of mode $m$. In the $(\Delta_m,G_m)$ plane, the gain window bounds the instability tongue of mode $m$.
Above threshold, an infinitesimal fluctuation grows exponentially, the Kerr frequency shift saturates this growth at a finite amplitude, and the mode thereby \emph{activates}.
Writing $\alpha_m=\sqrt{I_m}e^{i\phi_m}$, with intensity $I_m=|\alpha_m|^2$, the stationary solutions are
\begin{equation}
    \begin{aligned} I_{m,\sigma} &= \frac{\Delta_m - \sigma\chi_m}{U_m}, \\ \phi_{m,\sigma} &= -\frac12\arg\left(\sigma\chi_m + i\frac{\kappa_m}{2}\right) \;\mathrm{mod}\,\pi, \end{aligned}
    \label{eq:single_kpo_stationary_solutions}
\end{equation}
where $\sigma=\pm1$ labels the two response branches, and only solutions with $I_{m,\sigma}>0$ are physical. Each physical response branch has two phase states related by $\alpha_m\to-\alpha_m$.
A mode thus has one, three, or five FPs [Fig.~\ref{fig:1}(b)]. In the three-FP regime, the vacuum is a saddle whose unstable branches lead to the two stable phase states [Fig.~\ref{fig:1}(c)]. For $\Delta_m>\chi_m>0$, the vacuum is stable and coexists with two stable $\sigma=-1$ states and two $\sigma=+1$ saddles.

\section{Avalanches in a KPO dimer}
\label{sec:chains}

\begin{figure}[!t]
  \centering
  \includegraphics[width=0.48\textwidth]{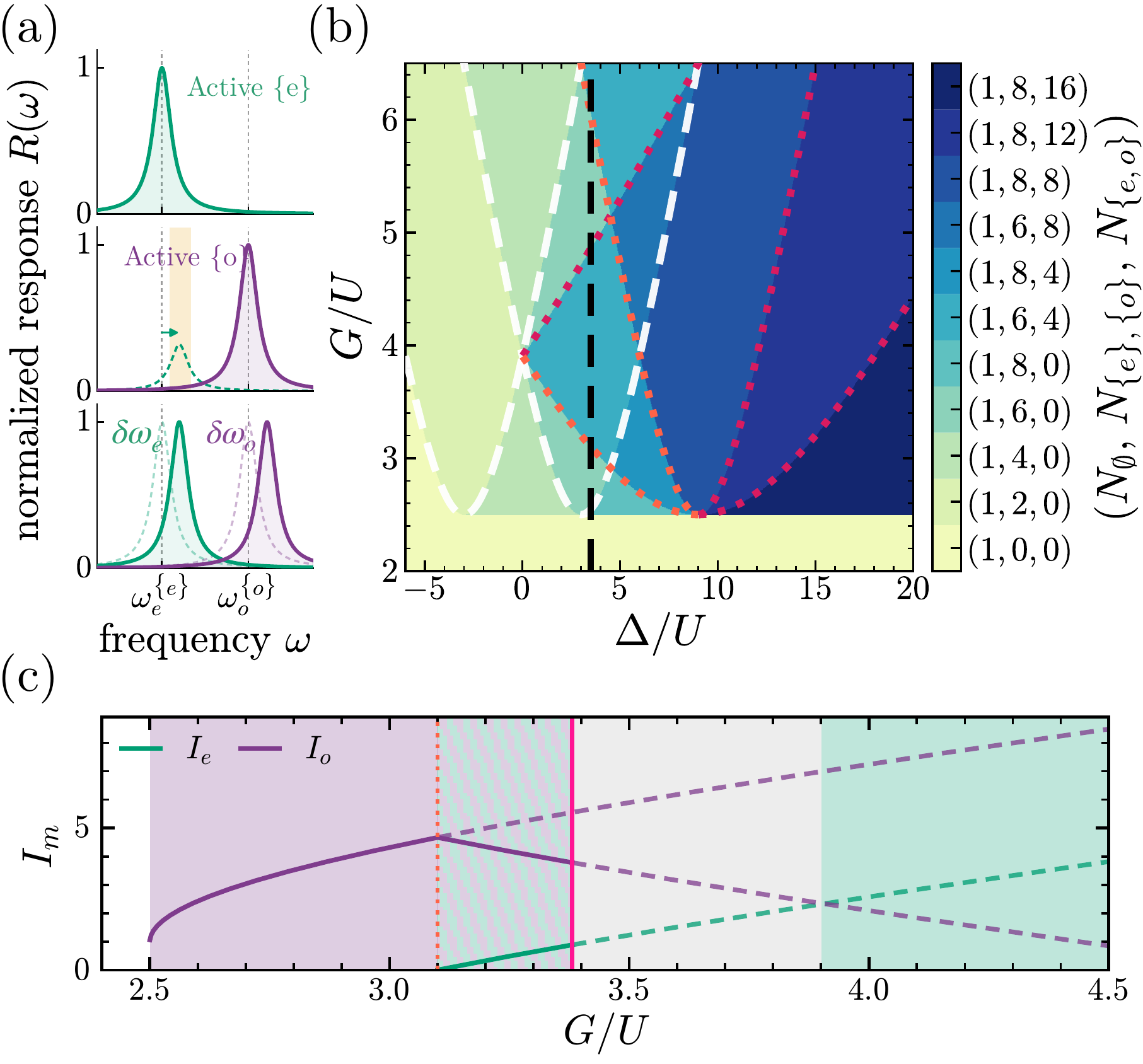}
  \caption{(a)~Schematic normalized modal responses: top, even mode active; middle, odd mode active, whose cross-Kerr shift carries the dormant even resonance (dashed) into the even-mode gain window (shaded); bottom, both modes active, with dashed monomode references. $R(\omega)$ is the normalized response of each mode, $\omega_m^{\{m\}}$ its resonance on the monomode $\{m\}$, and $\delta\omega_{e,o}$ the cross-Kerr shifts.
(b)~Stationary-state count in the $(\Delta/U,G/U)$ plane. Color triplets count empty-, single-, and two-mode FPs, respectively, including both stable and unstable states. White dashed curves, linear instability tongues $|\Delta_e|<\chi$ and $|\Delta_o|<\chi$; colored dotted curves, mixed branch boundaries; black dashed line, the $\Delta/U=3.5$ cut used in Fig.~\ref{fig3}.
(c)~Stationary intensities along the cut $\Delta/U=3.5$ of (b), with stable branches solid and unstable branches dashed; dotted line, mixed branch boundary as in (b); magenta line, Hopf point of the two-mode FP. Backgrounds denote the support reached from the vacuum along the cut, colored as in Fig.~\ref{fig3}(b).
We use $U$ as the unit of rates ($U=1$). Parameters: $\kappa/U=5$ and $J/U=3$.}
  \label{fig2}
\end{figure}
\begin{figure*}[t]
  \centering
  \includegraphics[width=\textwidth]{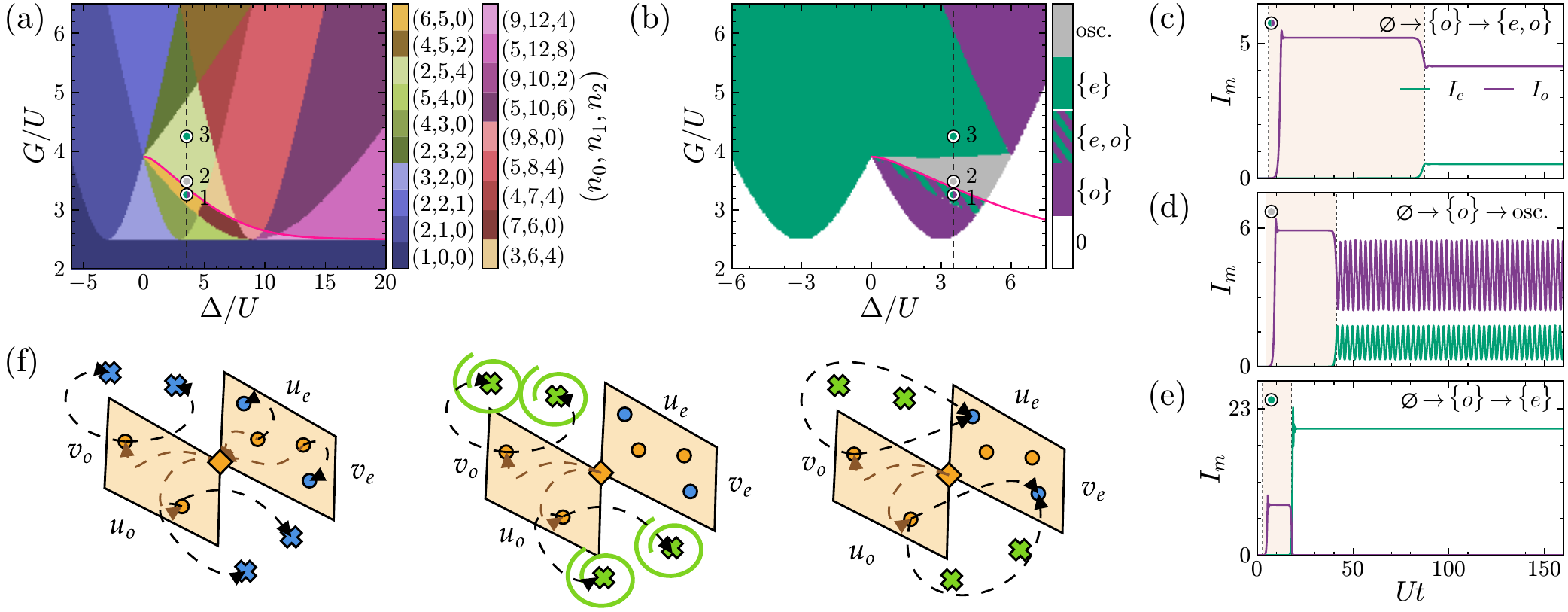}
  \caption{(a)~Morse-resolved FP landscape, labeled by the numbers $(n_0,n_1,n_2)$ of FPs with $0$, $1$, and $2$ unstable directions; magenta curve, Hopf boundary of the two-mode FP.
  (b)~Long-time states reached from vacuum perturbations, with the dashed cut and colored markers identifying the trajectories in panels (c)--(e).
  (c)~Two-mode avalanche, $\varnothing\to\{o\}\to\{e,o\}$.
  (d)~Capture by the limit cycle, $\varnothing\to\{o\}\to\mathrm{osc.}$
  (e)~Transfer to the even monomode, $\varnothing\to\{o\}\to\{e\}$. In (c)--(e), shading marks the odd-mode plateau and dotted lines the activations that end the vacuum dwell and the plateau.
  (f)~The four invariant subspaces of the four-dimensional phase space, one per support, and representative connections, with FPs colored by Morse index, for the three regimes of (c)--(e). The rectangles represent the invariant odd (left) and even (right) subspaces, crosses are mixed FPs, and the dashed connections leave the odd monomode saddles toward stable two-mode FPs, limit cycles around unstable two-mode FPs, or the even monomode, respectively.
  Panels (c)--(e) use the cut of (b), $\Delta/U=3.5$, with $G/U=3.26$, $3.49$, and $4.25$.
Parameters: $\kappa/U=5$, $J/U=3$, and $\varepsilon_0=10^{-3}$ in (b)--(e).}
  \label{fig3}
\end{figure*}

The cross-Kerr term in Eq.~\eqref{eq:general_hamiltonian} commutes with every $\hat{n}_m$, so it exchanges no excitations between modes. Instead, a populated mode shifts the detuning of its neighbors, bringing them closer to or further from their parametric threshold.
We call a mode $m$ \emph{active} when populated, $I_m>0$, and \emph{dormant} when $I_m=0$, and the set of active modes the \emph{support}. These labels describe the stationary states visited; nearby trajectories can retain small nonzero components in their dormant modes.
The active modes shift the effective detuning of a dormant mode $d$ to
\begin{equation}
    \Delta_d^{\mathrm{eff}}(\boldsymbol\alpha) = \Delta_d - \sum_{a}V_{d,a}I_a,
    \label{eq:effective_detuning}
\end{equation}
where the sum runs over the active modes $a$.
When this shift brings $\Delta_d^{\mathrm{eff}}$ into the gain window $|\Delta_d^{\mathrm{eff}}|<\chi_d$, a small nonzero amplitude in mode $d$ grows exponentially. As analyzed below, while this amplitude remains small the network lingers near the stationary state on the preceding support, before rapid saturation activates the new mode [Fig.~\ref{fig:1}(d)].
We call such a sequence of instability-driven support changes a \emph{modal avalanche}.

\paragraph*{Stationary activity patterns.}
To illustrate the mechanism, we consider two modes, $e$ and $o$ (even and odd), with detunings $\Delta_e=\Delta+J$ and $\Delta_o=\Delta-J$, self-Kerr coefficients $U/2$, and cross-Kerr couplings $V_{e,o}=V_{o,e}=U$, as for the normal modes of a two-junction array (Appendix~\ref{app:model_validity}). We assume both modes have the same drive $G$ and loss $\kappa$, so $\chi_e=\chi_o\equiv\chi$. The detuning split offsets their gain windows. When only one mode is unstable from the vacuum, its population can shift the other into its gain window and activate it [Fig.~\ref{fig2}(a)].

The four supports are $\varnothing$, $\{e\}$, $\{o\}$, and $\{e,o\}$. For a fixed nonempty support and response-branch signs, the stationary intensities follow from a linear system; only solutions with real $\chi$ and positive active intensities are physical (Appendix~\ref{app:dimer}). Different branches and phase states can therefore give several FPs on the same support.

Figure~\ref{fig2}(b) counts these existing FPs, irrespective of their stability or dynamical accessibility. Outside both instability tongues the vacuum is stable, although finite-amplitude states can coexist with it; entering a tongue destabilizes the vacuum along the corresponding mode. The colored dotted curves delimit regions with physical mixed-support branches.
Figure~\ref{fig2}(c) shows the intensities and stability of these branches along the cut $\Delta/U=3.5$, which we use below to study the trajectories.
\paragraph*{Conditional mode activation.}
The existence of a stationary branch does not ensure its stability against excitation of a dormant mode.
On the odd-mode FP, the intensity $I_o$ shifts the dormant even mode to $\Delta_e^{\rm eff}=\Delta_e-UI_o$. The dormant amplitude vanishes on the subspace of the support, which is invariant because Eq.~\eqref{eq:mf_general_complex} has no source term independent of a mode amplitude.
The even mode feels the odd one only through $\Delta_e^{\rm eff}$, so its block of the Jacobian is $\mathcal J_e^{(\varnothing)}$ with $\Delta_e$ replaced by $\Delta_e^{\rm eff}$; the full fluctuation Hamiltonian and linearization are given in Appendix~\ref{app:supports}. Its eigenvalues are \emph{transverse} because they govern perturbations outside the odd-mode subspace:
\begin{equation}
    \lambda_\pm^e(\{o\})
    =
    -\frac{\kappa_e}{2}
    \pm
    \sqrt{
        G_e^2-
        \left(\Delta_e-V_{e,o}I_o\right)^2
    },
    \label{eq:dimer_transverse_eigs}
\end{equation}
For $|\Delta_e^{\rm eff}|<\chi$, $\lambda_+^e(\{o\})>0$, so a small nonzero even-mode amplitude is amplified and the trajectory leaves the neighborhood of the invariant odd subspace. Increasing $I_o$ thus shifts the dormant even mode into its gain window [Fig.~\ref{fig2}(a)]. At $\Delta/U=3.5$ this transverse instability sets in above $G\simeq3.10\,U$ (Appendix~\ref{app:dimer}).
Instabilities within a support, which we call internal, instead change the dynamics without changing the support.

\paragraph*{Even--odd modal avalanches.}
Figure~\ref{fig3}(a) resolves the stationary-state counts of Fig.~\ref{fig2}(b) by Morse index: $n_0$ counts stable FPs and $n_1,n_2$ saddles.
Figure~\ref{fig3}(b) instead records the outcomes reached from vacuum perturbations: the vacuum, monomodes, mixed supports, or persistent activity, i.e., sustained motion that does not settle on an FP. The same Morse vector can yield different dynamical outcomes, as the examples below show (Appendix~\ref{app:dimer}).

We initialize the dimer at $\boldsymbol\alpha(0)=\varepsilon_0\hat{\boldsymbol e}$, a distance $0<\varepsilon_0\ll1$ from the vacuum along a unit direction $\hat{\boldsymbol e}$, at $\Delta/U=3.5$ and $G/U=3.26$, $3.49$, or $4.25$, inside the odd instability tongue and above the even-mode activation threshold [Fig.~\ref{fig3}(b)].
The vacuum is then a saddle whose only unstable direction is the odd mode, so the odd component grows exponentially while the even one decays, and the dimer rings up onto the odd monomode [Fig.~\ref{fig3}(c)--(e)].
The odd-monomode FP is a saddle with one transverse unstable direction. The trajectory therefore dwells near it, on a modal plateau, until the small even component left over from the vacuum dwell is amplified and activates the even mode.

The activation that ends this plateau depends on the drive, and so does the \emph{avalanche word}, the time-ordered sequence of visited supports.
At $G/U=3.26$ the dimer settles on the mixed FP of $\{e,o\}$, with both modes active [Fig.~\ref{fig3}(c)].
At $G/U=3.49$ the mixed FP has lost internal stability in a Hopf bifurcation at $G_H/U\simeq3.38$, and the flow is captured by a limit cycle, a persistent two-mode oscillation [Fig.~\ref{fig3}(d); Appendix~\ref{app:dimer}], similar to the persistent beating of linearly coupled parametric oscillators~\cite{bello2019persistent,calvanesestrinati2019theory}.
At $G/U=4.25$ the activation completes a transfer to the even monomode, $\varnothing\to\{o\}\to\{e\}$, the most elementary modal avalanche, in which one monomode hands its activity to the other [Fig.~\ref{fig3}(e)].
In phase space, each support spans an invariant subspace, and a trajectory can leave the neighborhood of a support only through a transverse instability, which enlarges it; for the odd monomode this corresponds to $\lambda_+^e(\{o\})>0$ [Eq.~\eqref{eq:dimer_transverse_eigs}]. The dimer flow is thus the minimal network of supports linked by transverse instabilities, and each outcome follows one of its links [Fig.~\ref{fig3}(f)].

\begin{figure}[!t]
  \centering
  \includegraphics[width=\columnwidth]{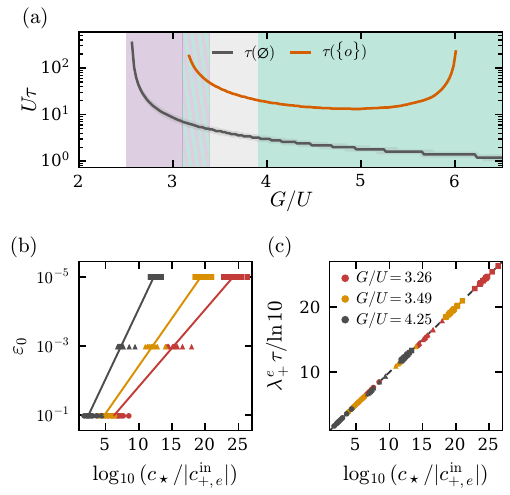}
  \caption{(a)~Dwell times in the vacuum, $\tau(\varnothing)$ (gray), and on the odd-mode plateau, $\tau(\{o\})$ (orange), along the cut $\Delta/U=3.5$ at $\varepsilon_0=10^{-3}$: medians (lines) and interquartile ranges (bands). Backgrounds denote the support reached along the cut, colored as in Fig.~\ref{fig3}(b).
(b)~Depth $\log_{10}[c_\star/|c_{+,e}^{\mathrm{in}}|]$ of the incoming seed on entering the odd-monomode neighborhood, for sixteen initial directions at $\varepsilon_0=10^{-1},10^{-3},10^{-5}$ and $G/U=3.26,3.49,4.25$.
Solid lines: seed-depth law with exponent $1+\gamma^e(\varnothing)/\lambda_+^o(\varnothing)$, the growth of the depth per decade of $\varepsilon_0$, anchored at the median for $\varepsilon_0=10^{-1}$.
(c)~Odd-monomode dwell times in local growth decades, $\lambda_+^e(\{o\})\tau(\{o\})/\ln 10$, versus incoming-seed depth;
dashed line, the prediction $\lambda_+^e(\{o\})\tau(\{o\})/\ln 10=\log_{10}[c_\star/|c_{+,e}^{\mathrm{in}}|]$. Colors: $G/U$. Marker shapes: $\varepsilon_0$.
Parameters: $\Delta/U=3.5$, $J/U=3$, and $\kappa/U=5$.}
  \label{fig4}
\end{figure}
\paragraph*{History-dependent dwell times.}
The dimer realizes one of three avalanche words,
\begin{equation}
\begin{tikzpicture}[baseline=(O.base),>=stealth,
    tau/.style={label={[inner sep=0pt,label distance=1pt]above:$\scriptstyle #1$}}]
  \node[tau={\tau(\varnothing)}] (V) at (0,0)   {$\varnothing$};
  \node[tau={\tau(\{o\})}]       (O) at (1.5,0) {$\{o\}$};
  \node[anchor=west] (EO)  at (2.9,0.5)  {$\{e,o\}$};
  \node[anchor=west] (OSC) at (2.9,0)    {$\mathrm{osc.}$};
  \node[anchor=west] (E)   at (2.9,-0.5) {$\{e\}$};
  \draw[->] (V) -- (O);
  \draw[->] (O.east) to[bend left=20]  (EO.west);
  \draw[->] (O.east) -- (OSC.west);
  \draw[->] (O.east) to[bend right=20] (E.west);
\end{tikzpicture}
\label{eq:dimer_word_dwells}
\end{equation}
where $\tau(S)$ denotes the dwell time on support $S$, spent near its saddle while the next mode grows from its seed, and the arrows represent the fast saturation of that mode. All three words pass through a vacuum dwell and an odd-mode plateau before reaching different outcomes.

Figure~\ref{fig4}(a) shows the computed dwell times on the two transient supports, the vacuum and the odd monomode, along the cut of Fig.~\ref{fig2}(c), with the shading marking the final state reached. The vacuum dwell $\tau(\varnothing)$ diverges at the vacuum threshold of the odd mode, and $\tau(\{o\})$ at both edges of the drive range in which the odd monomode is transversely unstable. At these bifurcations the gain vanishes, and the dwells show critical slowing down.

A \emph{seed}, a small amplitude $\varepsilon$ left by the initial perturbation or by preceding dynamics, grows as $\varepsilon e^{\lambda t}$ with the gain $\lambda$ of the unstable direction while the local linearization holds. Reaching an exit amplitude $c$ gives the dwell time
\begin{equation}
\tau(\varnothing)=\frac{1}{\lambda_+^m(\varnothing)}\ln\!\left(\frac{c}{\varepsilon}\right).
\label{eq:ring_up_time}
\end{equation}
written here for the vacuum. Local growth is exponential, while the exit delay increases logarithmically as the seed decreases. All dwell times below are of this form, with the argument of $\tau$ naming the state being left.

Each dwell depends not only on the instability of the saddle being left but also on the history of the trajectory.
Near the vacuum, the growing odd projection scales as $\varepsilon_0e^{\lambda_+^o(\varnothing)t}$. Meanwhile, the even component rotates within a decay envelope $\varepsilon_0e^{-\gamma^e(\varnothing)t}$. For the three drives considered, $\gamma^e(\varnothing)=\kappa/2$. Both expressions omit prefactors set by the initial direction. Here $\gamma^d(S)=-\max_\pm\operatorname{Re}\lambda_\pm^d(S)$ is the decay rate of a stable dormant direction, the counterpart of the gain $\lambda_+^d(S)$ of an unstable one.
We measure the local dwell up to a fixed exit magnitude $c_\star$ of the growing projection, playing the role of $c$ in Eq.~\eqref{eq:ring_up_time}. This marks departure from the saddle neighborhood, before nonlinear saturation, giving $\tau(\varnothing)\simeq\ln(c_\star/\varepsilon_0)/\lambda_+^o(\varnothing)$ to leading logarithmic order for a fixed initial direction with nonzero unstable projection.
The trajectory then reaches the odd saddle with the \emph{incoming seed} $c_{+,e}^{\mathrm{in}}$ of the next activation, prepared by the dynamics itself. This is the displacement's projection onto the dual left eigenvector of the growing even channel, evaluated on entry to the odd-saddle neighborhood and normalized consistently with $c_\star$ (Appendix~\ref{app:dimer}). The vacuum dwell contributes a decay-envelope factor $e^{-\gamma^e(\varnothing)\tau(\varnothing)}$, giving the leading power-law envelope
\begin{equation}
|c_{+,e}^{\mathrm{in}}|\sim\varepsilon_0\left(\frac{\varepsilon_0}{c_\star}\right)^{\gamma^e(\varnothing)/\lambda_+^o(\varnothing)}\propto\varepsilon_0^{\,1+\gamma^e(\varnothing)/\lambda_+^o(\varnothing)}.
\label{eq:dimer_seed_size}
\end{equation}
The prefactor oscillates with $\ln\varepsilon_0$ because the even component rotates while it decays, modulating its projection onto the growing channel (Appendix~\ref{app:dimer}). With this seed and the transverse gain $\lambda_+^e(\{o\})$, Eq.~\eqref{eq:ring_up_time} gives the plateau dwell, measured from entry into the saddle neighborhood,
\begin{equation}
\tau(\{o\})\simeq\frac{1}{\lambda_+^e(\{o\})}\ln\!\left(\frac{c_\star}{|c_{+,e}^{\mathrm{in}}|}\right).
\label{eq:dimer_inherited_seed_summary}
\end{equation}
The seed's modulus sets this dwell, while its sign selects one of the two $\mathbb Z_2$-related even branches. A smaller initial perturbation thus gives a longer vacuum dwell and a smaller incoming seed, so the plateau duration is history dependent.
Integrating Eq.~\eqref{eq:mf_general_complex} for sixteen initial directions $\hat{\boldsymbol e}$ at $\varepsilon_0=10^{-1},10^{-3},10^{-5}$ and the three drives of Fig.~\ref{fig3}(c)--(e), we find that the computed seeds follow this envelope [Eq.~\eqref{eq:dimer_seed_size}], with slopes set by the vacuum rates [Fig.~\ref{fig4}(b)].
For the same runs, the computed plateau dwells collapse onto Eq.~\eqref{eq:dimer_inherited_seed_summary} for all drives, seed amplitudes, and initial directions [Fig.~\ref{fig4}(c)], using the spectral gain and the measured incoming projection. The comparison has no fitted parameter, because the dwell is computed up to the same $c_\star$ that enters the seed depth, so changing $c_\star$ within the linear regime moves every point along the diagonal. This collapse tests the exponential growth from the incoming seed, whereas Fig.~\ref{fig4}(b) tests how that seed inherits the vacuum dwell.
The same local dwell law, Eq.~\eqref{eq:dimer_inherited_seed_summary}, thus governs all three outcomes of Fig.~\ref{fig3}(c)--(e); the branch of Eq.~\eqref{eq:dimer_word_dwells} is selected only after escape, by the connections of Fig.~\ref{fig3}(f).

\section{Aging in many-mode avalanches}\label{sec:landscapes}\label{sec:many_mode_aging}

\begin{figure*}[!tp]
  \centering
  \includegraphics[width=\textwidth]{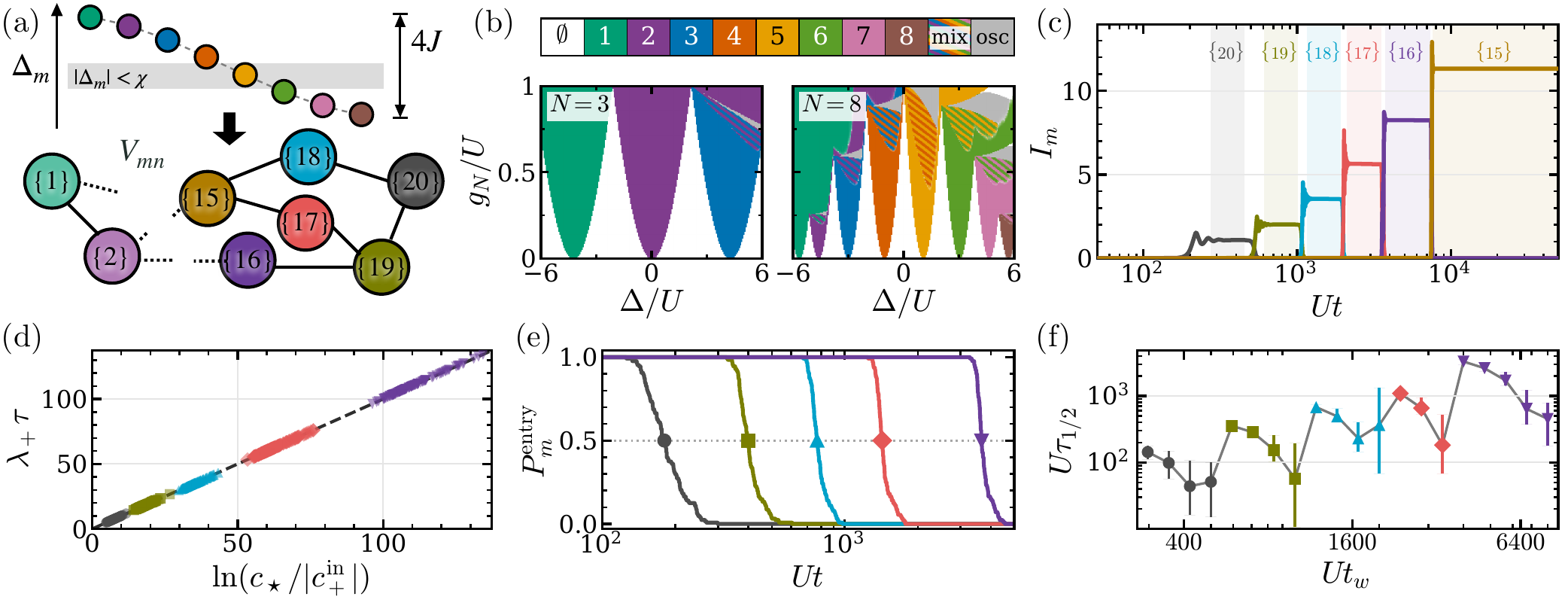}
  \caption{(a)~Top, detuning ladder of Eq.~\eqref{eq:detuning_ladder} for $N=8$, inside the interval $\Delta\pm2J$ of width $4J$, with the gain window $|\Delta_m|<\chi$ shaded; bottom, part of the modal network of the $N=20$ avalanche; nodes, single-mode supports $\{m\}$, with $\{15\}$--$\{20\}$ colored as in (c)--(f); links, modal couplings $V_{m,n}$ (all pairs are coupled; dotted lines lead to modes not shown).
(b)~Dominant finite-time outcomes for $N=3$ (left) and $N=8$ (right) in the plane of $\Delta/U$ and the normalized drive $g_N/U$ (0 at the loss threshold, 1 at the cutoff; Appendix~\ref{app:maps}), from 32 uniformly sampled initial directions per point. Key: $\varnothing$, vacuum; $m$, single-mode support $\{m\}$; mix, mixed supports (stripes); osc, persistent activity. Unresolved trajectories are colored by their endpoint support. Maps for $N=3$--$8$ are in Appendix~\ref{app:maps}. Parameters: $\kappa/U=5$, $J/U=3$, $\varepsilon_0=10^{-3}$, and $UT=5000$.
(c)~Modal intensities of the representative trajectory of the $N=20$ avalanche, closest to the five median log dwells; shading and labels, plateaus $\{20\}$--$\{15\}$.
(d)~Rescaled dwell $\lambda_+(\{m\})\tau(\{m\})$ versus incoming seed depth $\ln[c_\star/|c_+^{\mathrm{in}}(\{m\})|]$ for all 640 escapes; dashed line, equality.
(e)~Entry persistence $P_m^{\mathrm{entry}}(t)$ on each transient plateau for the 128 trajectories; symbols, median dwells.
(f)~Median remaining lifetime $\tau_{1/2}$ on the plateau occupied at the waiting time $t_w$; bars, 16th--84th percentiles; gray line, connected medians.
Plateau colors and markers are common to (c)--(f). Parameters in (c)--(f): $N=20$, $J/U=1$, $\Delta/U=1.99$, $\kappa/U=0.057$, $\chi/U=0.067$, $G/U=0.0728$, and $\varepsilon_0=10^{-3}$.}
  \label{fig:many_mode_aging}
\end{figure*}

\paragraph*{From the dimer to many-mode landscapes.}
Increasing $N$ enlarges the set of supports and transverse escape channels.
We consider $N$ normal modes of a uniform array, motivated by the extended modes of Josephson junction arrays~\cite{weissl2015kerr,krupko2018kerr}. Their detunings form the ladder
\begin{equation}
\Delta_m=\Delta+2J\cos k_m,\qquad k_m=\frac{\pi m}{N+1},
\label{eq:detuning_ladder}
\end{equation}
which lies within the band $\Delta\pm2J$ and reduces for $N=2$ to the dimer [Fig.~\ref{fig:many_mode_aging}(a)].
The normal-mode overlaps give self-Kerr coefficients $V_{m,m}=3U/[2(N+1)]$ and cross-Kerr couplings $V_{m,n}=2U/(N+1)$, with enhanced values for mirror pairs and, for odd $N$, the central mode (Appendix~\ref{app:model_validity}).
The lower network in Fig.~\ref{fig:many_mode_aging}(a) shows the cross-Kerr couplings: population in one mode shifts the detunings of the others and can activate them. Its links represent couplings, while the support transitions actually followed depend on the trajectory.
\paragraph*{Support-resolved stationary states and transverse activation.}
For a fixed support $S$ and branch signs, the stationary intensities follow from a linear system (Appendix~\ref{app:supports}). Their stability separates into internal motion on $S$ and transverse perturbations in dormant modes, generalizing the dimer analysis of Sec.~\ref{sec:chains}.
On an FP of support $S$, a dormant mode $d$ sees the effective detuning $\Delta_d^{\mathrm{eff},(S)}$ of Eq.~\eqref{eq:effective_detuning}, and its transverse eigenvalues take the single-mode form
\begin{equation}
\lambda_\pm^d(S)=
-\frac{\kappa}{2}\pm
\sqrt{G^2-\bigl(\Delta_d^{\mathrm{eff},(S)}\bigr)^2}.
\label{eq:transverse_eigs_general}
\end{equation}
We call the dormant modes with $\lambda_+^d(S)>0$ the \emph{activation set} of $S$, the channels through which the trajectory can leave.
A trajectory traces its word $\varnothing=S_0\to S_1\to S_2\to\cdots$ through the network of supports, dwelling on each and leaving through one channel of its activation set; which links it follows is a global question~\cite{ashwin1999heteroclinic,villa2025topological,gomez2025quantum}.
\paragraph*{Competing channels and incoming seeds.}
Each channel rings up from its own incoming seed $c_{+,d}^{\mathrm{in}}(S)$, the component of the arriving displacement along channel $d$. By Eq.~\eqref{eq:ring_up_time}, it reaches $c_\star$ after
\begin{equation}
\tau^d(S)=\max\!\left[0,
\frac{1}{\lambda_+^d(S)}
\ln\!\left(\frac{c_\star}
{|c_{+,d}^{\mathrm{in}}(S)|}\right)\right],
\label{eq:transverse_seed_race}
\end{equation}
where the maximum covers seeds that arrive at or above $c_\star$. Within this local linear description, the first crossing estimates the dwell, $\tau(S)\simeq\min_d\tau^d(S)$, and depends jointly on gain and incoming seed. The subsequent nonlinear flow selects the next support.
\paragraph*{Outcome maps from the vacuum.}
Figure~\ref{fig:many_mode_aging}(b) shows outcome maps for $N=3$ and $N=8$; the full set for $N=3$--$8$ is in Appendix~\ref{app:maps}. The normalized drive $g_N/U$ runs from the loss threshold at 0 to the closing of the last vacuum-stable gap between adjacent instability tongues at 1. At low drives, the maps resolve the individual vacuum-instability tongues. Increasing the drive takes the outcomes from monomodes through mixed supports to persistent activity, repeating the dimer sequence between neighboring tongues. These are finite-time outcomes: white near-vacuum endpoints need not represent a stable vacuum attractor (Appendix~\ref{app:maps}).

\paragraph*{Aging along an avalanche.}
To follow an avalanche one activation at a time, we now turn to a larger $N=20$ network with different parameters and smaller losses, $J/U=1$ and $\kappa/U=0.057$. Its drive $g_{20}/U\simeq0.36$ lies in the monomode regime (Appendix~\ref{app:maps}). Figure~\ref{fig:many_mode_aging}(c) follows the resulting avalanche from mode $20$ to mode $15$, starting, as in the dimer, from a small perturbation $\varepsilon_0\hat{\boldsymbol e}$ of the vacuum. Its six plateaus comprise five transient dwells, each ending in an escape, and a final stable plateau.
The activity passes from mode $20$ down the index, one mode per plateau, but up in detuning. A larger detuning requires a larger Kerr shift and hence a larger stationary intensity, as in the single-mode branch of Sec.~\ref{sec:model}.
The five transient plateaus last progressively longer, so the activations spread out in time and the avalanche slows down even as the network keeps amplifying.
Here \textit{aging} denotes the systematic growth of dwell times along the avalanche; each individual escape remains exponential.
This slowing connects to aging in glasses~\cite{bouchaud1998outofequilibrium,berthier2011theoretical}, active glasses~\cite{mandal2020multiple,janzen2022aging}, intermittent elastic relaxation~\cite{shohat2023logarithmic,korchinski2025thermal,shohat2025emergent}, and glassy states of coupled-oscillator ensembles~\cite{iatsenko2014glassy,choe2016incoherent}, where disorder and microscopic rearrangements are often intertwined~\cite{coniglio2000jamming,zaitsev2014slow,lahini2017nonmonotonic,dillavou2018nonmonotonic,rudyak2025spatiotemporal}. Here the slowing arises without quenched disorder.

Each dormant mode waits near zero amplitude until its turn, and meanwhile its seed decays. Mode $18$, for instance, decays during the vacuum dwell and the plateau on mode $20$, and grows only once mode $19$ holds the plateau. Later modes therefore start from smaller seeds and take longer to ring up, which repeats the single step of the dimer at every plateau.
We label each plateau by its monomode support $\{m\}$, whose single escape channel is mode $m-1$, with gain $\lambda_+(\{m\})$ and incoming seed $c_+^{\mathrm{in}}(\{m\})$.
For 128 initial directions at fixed $\varepsilon_0$, every trajectory follows the same six-plateau word. Repeating the dimer test of Fig.~\ref{fig4}(c) on every transient plateau, Fig.~\ref{fig:many_mode_aging}(d) compares the measured dwell in growth times with the incoming seed depth. All $640$ escapes fall on the diagonal predicted by Eq.~\eqref{eq:transverse_seed_race}, within $0.045$ growth times and with no free parameter, and the plateaus line up along it in the order of the word.
Each escape is thus an ordinary exponential ring-up, and the slowing comes from ever deeper incoming seeds together with the plateau-dependent gains.

\paragraph*{Dwell growth.}
Having tested each escape, we ask whether dwell growth persists across different initial directions. We align each trajectory with its own entry into plateau $\{m\}$ and measure the fraction that has not yet left after a time $t$, $P_m^{\mathrm{entry}}(t)=\Pr[\tau(\{m\})>t]$. These curves drop steeply on each plateau and move to longer times from $m=20$ to $m=16$, with no rescaling [Fig.~\ref{fig:many_mode_aging}(e)].
In every trajectory the five dwells increase, and their medians roughly double, with a geometric fit giving $q=2.075$ for this finite avalanche (Appendix~\ref{app:deterministic_methods}).
Geometric dwell growth spaces the activations nearly equally in $\ln t$, with an event count $k(t)\sim\ln t/\ln q$ (Appendix~\ref{app:deterministic_methods}). Logarithmic relaxation is common in complex systems~\cite{amir2012relaxations,lomholt2013microscopic}, and in spin glasses and crumpled sheets events occur at a constant average rate in $\ln t$~\cite{sibani2003logpoisson,lahini2023crackling,boettcher2023instability}. Here the spacing is deterministic.

\paragraph*{Waiting-time persistence.}
Aging studies of glasses let a system age for a waiting time $t_w$ after its preparation and then record its relaxation over a further time $t$~\cite{bouchaud1998outofequilibrium,berthier2011theoretical}.
Following this protocol, we define the waiting-time persistence $P(t_w,t)$ as the probability that a trajectory sitting on a transient plateau at $t_w$ stays there throughout the next interval $t$.
\begin{equation}
\begin{aligned}
P(t_w,t)=\Pr\bigl[&\tau(\{m\})>t_w-t_{\mathrm{in}}(\{m\})+t\\
&\,\bigm|\,\tau(\{m\})>t_w-t_{\mathrm{in}}(\{m\})\bigr],
\end{aligned}
\label{eq:global_support_persistence}
\end{equation}
where $\{m\}$ is the transient plateau occupied at $t_w$ and $t_{\mathrm{in}}(\{m\})$ is its entry time. Unlike the entry-aligned clock in Fig.~\ref{fig:many_mode_aging}(e), $t_w$ is measured from the initial preparation. The median remaining lifetime $\tau_{1/2}(t_w)$ satisfies $P(t_w,\tau_{1/2})=1/2$; it decreases as the trajectory approaches its next activation and rises on entering the next, slower plateau, producing a sawtooth [Fig.~\ref{fig:many_mode_aging}(f)].
The phenomenological form~\cite{phillips1996stretched,bouchaud1998outofequilibrium}
\begin{equation}
P(t_w,t)\simeq\exp\!\left[-\left(\frac{t}{z_0t_w^{\mu_P}}\right)^\beta\right],
\label{eq:global_kww_fit}
\end{equation}
with scale $z_0$, shape $\beta$, and aging exponent $\mu_P$, gives $\mu_P\simeq0.98$ when pooling observations from successive plateaus. This is close to the full aging $\mu_P=1$ of many glasses~\cite{bouchaud1992weak}, but does not imply that a single plateau slows internally, and the curves for different $t_w$ do not collapse onto a single function of $t/t_w$ (Appendix~\ref{app:numerical_methods}).

\paragraph*{Accumulated seed decay.}
Along the word $S_0=\varnothing,S_1,\ldots,S_K$, let $d$ be the mode that ends plateau $S_k$. Its seed starts at order $\varepsilon_0$ and decays with an envelope rate $\gamma^d(t)$. The evolving modal intensities change its effective detuning through cross-Kerr shifts, and hence its transverse dynamics. Equation~\eqref{eq:ring_up_time} then gives
\begin{equation}
\tau(S_k)\simeq\frac{1}{\lambda_+^d(S_k)}\Bigl[\ln\frac{c_\star}{\varepsilon_0}+\int_0^{t_{\mathrm{in}}(S_k)}\gamma^d(t)\,dt\Bigr],
\label{eq:general_inherited_dwell_law}
\end{equation}
where $t_{\mathrm{in}}(S_k)$ is the arrival time at $S_k$.
In this avalanche, every future escape mode has a complex pair of transverse eigenvalues on every earlier support, so its seed has a decay envelope $\gamma^d=\kappa/2$. Assuming the same rate during the support changes, the integral becomes $(\kappa/2)\,t_{\mathrm{in}}(S_k)$ (Appendix~\ref{app:deterministic_methods}). A plateau reached later therefore holds the trajectory longer.

\paragraph*{Spectral prediction of the preparation dependence.}
Equation~\eqref{eq:general_inherited_dwell_law} predicts how the dwells change with $\varepsilon_0$. If support changes and transfer factors add no leading term in $\ln(1/\varepsilon_0)$, the dwell envelopes are linear in this logarithm. Differentiating Eq.~\eqref{eq:general_inherited_dwell_law}, with $S_1=\{20\},\ldots,S_5=\{16\}$, gives
\begin{equation}
\frac{\partial\tau(S_k)}{\partial\ln(1/\varepsilon_0)}
\simeq\frac{1}{\lambda_+(S_k)}\Bigl(1+\frac{\kappa}{2\lambda_+^{20}(\varnothing)}\Bigr)
\prod_{j=1}^{k-1}\Bigl(1+\frac{\kappa}{2\lambda_+(S_j)}\Bigr),
\label{eq:spectral_delay_product}
\end{equation}
where each earlier support contributes a factor set by the ratio of decay to transverse gain.
Varying $\varepsilon_0$ from $10^{-2}$ to $10^{-6}$, all 128 directions keep the six-plateau word, and the median dwells are linear in $\log_{10}\varepsilon_0$ ($R^2=0.992$--$0.998$), with a delay per decade that grows about tenfold from plateau $\{20\}$ to $\{16\}$, and Eq.~\eqref{eq:spectral_delay_product}, which has no fitted parameter, reproduces every slope within $1.6\%$ (Appendix~\ref{app:deterministic_methods}).
The timing of the avalanche thus remembers the initial amplitude, each plateau more strongly than the last.

\paragraph*{Beyond simple avalanche words.}
\begin{figure}[!tp]
  \centering
  \includegraphics[width=\columnwidth]{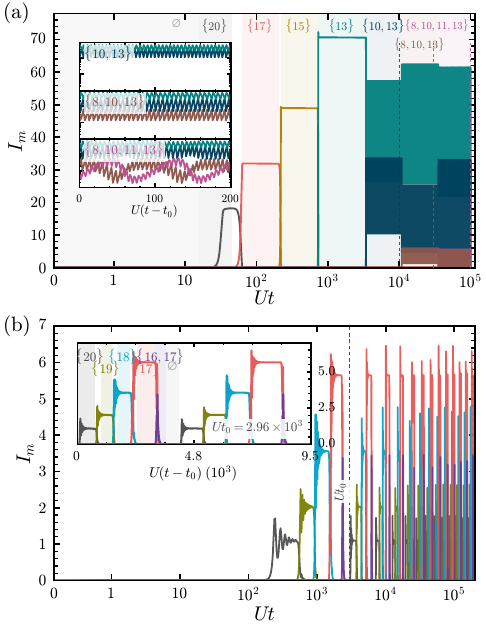}
  \caption{Modal intensities $I_m$ for two trajectories starting near the vacuum, with the visited supports labeled.
(a)~Insets: late-time motion on the mixed supports; dashed lines, inset time origins.
(b)~Representative trajectory. Inset: two early loops on a linear time axis, the first one shaded.
Parameters: $N=20$ and $\varepsilon_0=10^{-3}$; (a)~$(J,\Delta,\kappa,G)/U=(3,6,5,2.79)$, $U\,dt=0.05$, and $UT=10^5$; (b)~$(J,\Delta,\chi,\kappa,G)/U=(1,2,0.056,0.028,0.0577)$, $U\,dt=0.02$, and $UT=2\times10^5$.}
  \label{fig:complex_cascades}
\end{figure}
The six-plateau avalanche is the simplest one, in which each support is visited once, in a fixed order, and the trajectory ends on a stationary state.
Figure~\ref{fig:complex_cascades} shows two richer avalanches, one that ends in persistent oscillation on a mixed support and a cyclic avalanche, whose word returns to an earlier support and repeats.
Figure~\ref{fig:complex_cascades}(a) shows a network with $J/U=3$ and $\kappa/U=5$, as in the dimer, driven well above the cutoff $G_{\max}^{(N)}$ of Eq.~\eqref{eq:gmax_adjacent_crossing}. Four modes near the edge of the ladder, $m=17$--$20$, are then unstable from the vacuum at once.
Mode $20$ grows first, but the trajectory settles on $\{17\}$. The subsequent word skips modes, $\{17\}\to\{15\}\to\{13\}$.
Later activations lead to mixed supports, and the trajectory ends on $\{8,10,11,13\}$. There it oscillates at two resolved frequencies (Appendix~\ref{app:numerical_methods}), a many-mode counterpart of the dimer limit cycle.

In Fig.~\ref{fig:complex_cascades}(b), the drive is instead back in the monomode regime, close to that of Fig.~\ref{fig:many_mode_aging} but with half the loss.
The word starts as in Fig.~\ref{fig:many_mode_aging}, $\{20\}\to\{19\}\to\{18\}\to\{17\}$, but after a transient mixed-support episode on $\{16,17\}$, the trajectory falls back close to the vacuum and mode $20$ reactivates. All eight sampled initial directions repeat this loop.
The avalanche thus becomes cyclic, looping through the vacuum. This cyclic avalanche itself ages, its median period growing $7.5$-fold, from $3.80\times10^3\,U^{-1}$ to $2.85\times10^4\,U^{-1}$, over the first eleven loops.
The loop is a heteroclinic cycle through five saddles, $\varnothing\to\{20\}\to\{19\}\to\{18\}\to\{17\}\to\varnothing$, of the kind whose passages slow down geometrically in systems with invariant subspaces~\cite{may1975nonlinear,guckenheimer1988structurally,krupa1995asymptotic,krupa1997robust} (Appendix~\ref{app:deterministic_methods}).

\section{Discussion \& Outlook}\label{sec:discussion}
We have shown that parametrically driven Kerr networks can develop aging modal avalanches as parametric amplification drives them toward a nonequilibrium steady state.
Parametric gain first activates the modes whose detuning lies within their gain window, and their cross-Kerr shifts then pull dormant modes into theirs, triggering successive activations.
Dormant modes decay while they remain outside their gain windows, so later activations inherit smaller seeds. These seeds and the local gains produce nearly geometric dwell times and approximately logarithmic event accumulation, without disorder or energy barriers, and cyclic avalanches age from loop to loop.
The stationary states on each support and their stability follow analytically, whereas the avalanche word is set by the global connections of the network.
The inherited seeds also store memory of the initial amplitude: for $N=20$, the loss and transverse gain rates predict median dwell increases per decade reduction in initial amplitude within 1.6\%. This growth of successive dwell times requires dormant seeds to keep decaying, so seeds injected by residual parity-breaking forcing must remain below the inherited ones (Appendices~\ref{app:model_validity} and~\ref{app:numerical_methods}).
These results extend the phenomenology of sequential switching~\cite{ashwin1999heteroclinic,rabinovich2006dynamical,rabinovich2011robust,ashwin2007dynamics}, avalanches~\cite{sethna2001crackling,shohat2025emergent}, and aging~\cite{bouchaud1998outofequilibrium} to parametrically driven nonlinear oscillator networks. Here, parametric gain sustains the cascade, while dwell times depend on local amplification and seeds inherited from earlier motion.

A key ingredient of modal avalanches is already available in a chain of 21 superconducting resonators, where the occupation of one pumped normal mode shifts the frequencies of all the others~\cite{castillomoreno2026multimode}.
Avalanche transients may set how fast oscillator networks for optimization and computing~\cite{nikonov2015coupled,marandi2014network,inagaki2016large,goto2016bifurcation,yamamoto2017coherent,leleu2019destabilization} reach the configuration that is read out as a solution, and nonreciprocal couplings~\cite{fruchart2021nonreciprocal,delpino2022nonhermitian} or topological bands~\cite{peano2016topological,ozawa2019topological} could bias which avalanche words are realized.
These results suggest controlling avalanche timing through preparation and its sequence through engineered couplings.

\section*{Data availability}
The data that support the findings of this article, together with the code used to generate them and the figures, will be made openly available in a public repository upon publication and are available from the authors upon reasonable request.

\begin{acknowledgments}
We acknowledge funding from the Ram\'on y Cajal program (RYC2023-043827-I), funded by MICIU/AEI (10.13039/501100011033) and FSE+, and from the ``Mar\'ia de Maeztu'' Programme for Units of Excellence in R\&D (CEX2023-001316-M), funded by the Spanish Ministry of Science, Innovation and Universities.
J.d.P. also acknowledges support from the Proyectos de Generaci\'on de Conocimiento program, Grant No.~PID2024-158923NA-I00, funded by MICIU/AEI (10.13039/501100011033) and FEDER, UE.
A.S.G. acknowledges support from the predoctoral contract PREX2023-000117, funded by MCIN/AEI/10.13039/501100011033.
\end{acknowledgments}

\appendix
\section{\texorpdfstring{Circuit realization and parameters of the effective network}{Circuit realization and parameters of the effective network}}
\label{app:model_validity}

This appendix derives the circuit Kerr coefficients and a bound on $V_{m,n}$, the modal coupling matrix used in Secs.~\ref{sec:chains} and~\ref{sec:landscapes}, and the cutoff $G_{\max}^{(N)}$ of Appendix~\ref{app:maps}. It also states the conditions under which the effective network applies and defines the occupation scale.

\subsection{\texorpdfstring{Circuit Hamiltonian and Kerr coefficients}{Circuit Hamiltonian and Kerr coefficients}}
\label{app:circuit_kerr}
We consider $N$ normal modes of frequencies $\omega_{0,m}$ coupled to Josephson junctions $\ell$ with energies $E_\ell>0$.
The phase across junction $\ell$ is $\hat\varphi_\ell=\sum_m\varphi_{\ell,m}(\hat a_m+\hat a_m^\dagger)$, where the real coefficients $\varphi_{\ell,m}$ are the zero-point phase participations~\cite{blais2021circuit}.
For small junction phases, we expand the Josephson potential to quartic order.
After normal ordering, absorbing the quadratic terms into the linear modes, and dropping an additive constant, the expansion gives
\begin{equation}
 \hat H_{\rm circ}=\sum_m\omega_{0,m}\hat n_m-\sum_\ell\frac{E_\ell}{24}:\hat\varphi_\ell^4:+\hat H_d(t),
\label{eq:appC_circuit_hamiltonian}
\end{equation}
where colons denote normal ordering and $\hat H_d(t)$ is the drive Hamiltonian.
We set $\hbar=1$, so $E_\ell$ is in frequency units and $\omega_{0,m}$ includes the normal-ordering shifts.
Each mode is pumped near its own two-photon resonance, $\omega_{p,m}\simeq2\omega_{0,m}$, by multitone flux pumping or individual drives~\cite{grimm2020stabilization,sandbochang2018generating,chono2022twoqubit}.
In the frame rotating at $\omega_{p,m}/2$ for each mode, the quadratic term is $-\Delta_m\hat n_m$, with $\Delta_m=\omega_{p,m}/2-\omega_{0,m}$, and the resonant pump gives $G_m(\hat a_m^{\dagger2}+\hat a_m^2)/2$ after choosing the phase of the mode.
The quartic terms that conserve every modal occupation are also static.
Their six self-Kerr and twenty-four cross-Kerr orderings give the coefficients of Eq.~\eqref{eq:general_hamiltonian}~\cite{weissl2015kerr,krupko2018kerr,kuzmin2019superstrong,collodo2020implementation},
\begin{equation}
 \begin{aligned}
 U_m&=-\frac12\sum_\ell E_\ell\varphi_{\ell,m}^4,\\
 V_{m,n}&=-\sum_\ell E_\ell\varphi_{\ell,m}^2\varphi_{\ell,n}^2,\qquad m\ne n.
 \end{aligned}
\label{eq:appC_circuit_kerr}
\end{equation}
Appendix~\ref{app:rwa_parities} gives the rotating-wave conditions under which the other quartic and pump-induced terms can be dropped.

The Josephson coefficients in Eq.~\eqref{eq:appC_circuit_kerr} are negative, whereas the main text uses positive Kerr coefficients.
For the real drives used here, Eq.~\eqref{eq:mf_general_complex} is invariant under $\alpha_m\mapsto i\alpha_m^*$ together with $(\Delta_m,\mathbf V,G_m)\mapsto(-\Delta_m,-\mathbf V,G_m)$.
Thus changing the Kerr sign and reversing the detunings leaves the intensity dynamics unchanged, with the same positive drive convention.

\subsection{\texorpdfstring{Modal profiles and the nonlinear coupling matrix}{Modal profiles and the nonlinear coupling matrix}}
\label{app:modal_kerr_matrix}
The coupling matrix of Secs.~\ref{sec:chains} and~\ref{sec:landscapes} follows from a uniform-profile model with $N$ equivalent nonlinear elements, $E_\ell=E_J$, and a common phase participation scale $\varphi_0$:
\begin{equation}
 \begin{aligned}
 \varphi_{\ell,m}&=\varphi_0\psi_m(\ell),\\
 \psi_m(\ell)&=\sqrt{\frac{2}{N+1}}\sin\!\left(\frac{\pi m\ell}{N+1}\right),
 \qquad \ell,m=1,\ldots,N.
 \end{aligned}
\label{eq:appA_modal_profiles}
\end{equation}
These normalized sine profiles are those of a uniform open chain.
The common scale is an assumption of the model; mode-dependent phase participations in a particular circuit instead enter through Eq.~\eqref{eq:appC_circuit_kerr}.
In the positive-Kerr convention of Appendix~\ref{app:circuit_kerr}, we define the Kerr scale $U=E_J\varphi_0^4/2>0$.
The self- and cross-Kerr coefficients are then $U_m=U\sum_\ell\psi_m(\ell)^4$ and $V_{m,n}=2U\sum_\ell\psi_m(\ell)^2\psi_n(\ell)^2$ for $m\ne n$.
The sine sums give
\begin{equation}
 \begin{aligned}
 \sum_\ell\psi_m(\ell)^4&=\frac{3+\delta_{2m,N+1}}{2(N+1)},\\
 \sum_\ell\psi_m(\ell)^2\psi_n(\ell)^2&=\frac{2+\delta_{m+n,N+1}}{2(N+1)},\quad m\ne n,
 \end{aligned}
\label{eq:appA_modal_overlaps}
\end{equation}
so that the full matrix, with $V_{m,m}=U_m$, is
\begin{equation}
 \begin{aligned}
 \mathbf V&=\frac{U}{N+1}\boldsymbol\Gamma,\\
 \Gamma_{m,n}&=\begin{cases}
 \tfrac12(3+\delta_{2m,N+1}),&m=n,\\
 2+\delta_{m+n,N+1},&m\ne n.
 \end{cases}
 \end{aligned}
\label{eq:appA_modal_kerr_matrix}
\end{equation}
For a noncentral mode, a generic cross-Kerr coefficient is $4/3$ of its self-Kerr coefficient.
Mirror modes, $m+n=N+1$, have identical intensity profiles and the enhanced coefficient $3U/(N+1)$; for odd $N$, the central mode has self-Kerr coefficient $2U/(N+1)$.
We retain these mirror and central-mode corrections in all calculations.

More generally, applying Cauchy--Schwarz to the weighted vectors $\sqrt{E_\ell}\varphi_{\ell,m}^2$ in Eq.~\eqref{eq:appC_circuit_kerr} gives
\begin{equation}
 |V_{m,n}|\le2\sqrt{|U_mU_n|},\qquad m\ne n.
\label{eq:appA_kerr_bound}
\end{equation}
Equality holds when the two intensity profiles over the junctions are proportional, as for the mirror modes above.
For $N=2$, both profiles have $\psi_m(\ell)^2=1/2$, so $U_e=U_o=U/2$ and $V_{e,o}=U$, exactly the dimer of Sec.~\ref{sec:chains}.
The same construction gives the matrix used in Sec.~\ref{sec:landscapes}.

\subsection{\texorpdfstring{Rotating-wave approximation and modal parities}{Rotating-wave approximation and modal parities}}
\label{app:rwa_parities}
Write $\bar\omega_m=\omega_{p,m}/2=\omega_{0,m}+\Delta_m$ for the rotation frequency of mode $m$.
A static four-wave-mixing term $\hat a_{m_1}^\dagger\hat a_{m_2}^\dagger\hat a_{m_3}\hat a_{m_4}$ acquires the phase $e^{i\varpi_{m_1m_2m_3m_4}t}$, with
\begin{equation}
 \begin{aligned}
 \varpi_{m_1m_2m_3m_4}
 &=\bar\omega_{m_1}+\bar\omega_{m_2}-\bar\omega_{m_3}-\bar\omega_{m_4}\\
 &=\omega_{0,m_1}+\omega_{0,m_2}-\omega_{0,m_3}-\omega_{0,m_4}\\
 &\quad+\Delta_{m_1}+\Delta_{m_2}-\Delta_{m_3}-\Delta_{m_4}.
 \end{aligned}
\label{eq:appA_fwm_mismatch}
\end{equation}
Self- and cross-Kerr terms have $\varpi=0$ and are retained.
Other quartic terms rotate at the corresponding signed sums of $\bar\omega_m$, including $\bar\omega_{m_1}+\bar\omega_{m_2}+\bar\omega_{m_3}-\bar\omega_{m_4}$ for a three-to-one process.
The quadratic drive tone $\omega_{p,m'}$ of mode $m'$ creates a pair $(m,n)$ at mismatch $\bar\omega_m+\bar\omega_n-\omega_{p,m'}$, or converts modes at $\bar\omega_m-\bar\omega_n\pm\omega_{p,m'}$.
Only the intended pair $(m,n)=(m',m')$ is retained for each tone.

For every discarded term with nonzero coefficient, a sufficient secular condition is
\begin{equation}
 |\varpi|\gg\Omega,\qquad
 \Omega=\max_m\!\left(|\Delta_m|,G_m,\kappa_m,
 \sum_n|V_{m,n}|I_n\right),
\label{eq:appA_rwa_condition}
\end{equation}
with $\Omega$ also bounding the rates generated by the discarded couplings at the occupied amplitudes.
Averaging over times long compared with $|\varpi|^{-1}$ but short compared with $\Omega^{-1}$ then suppresses these rapidly oscillating terms.
Such effective models have been compared with full-circuit simulations for individually driven, well-separated KPOs~\cite{chono2022twoqubit}.
Flux pumping can also modulate the quartic interaction. The modulated monomials acquire additional frequency shifts $\pm\omega_{p,m'}$ and must satisfy the same separation, or have negligible amplitudes if resonant. Pump-induced frequency shifts must be calibrated or compensated~\cite{chono2025highperformance}.

The condition concerns all allowed frequency combinations, not just adjacent mode spacings.
For example, conversion between distinct mirror pairs, $\hat a_m^\dagger\hat a_{N+1-m}^\dagger\hat a_n\hat a_{N+1-n}$, is resonant whenever their rotation-frequency sums coincide.
A cosine bare spectrum has equal mirror-pair sums. The conversion mismatch in Eq.~\eqref{eq:appA_fwm_mismatch} then comes only from the small detunings and does not supply the separation of Eq.~\eqref{eq:appA_rwa_condition}.
The sine-profile Kerr matrix alone therefore does not justify discarding mode conversion. The bare spectrum and the pump tones must also avoid these resonances.
Individual pumps allow the effective ladder of Eq.~\eqref{eq:detuning_ladder} to be set separately from the bare frequencies, subject to these conditions.

The full quartic circuit and its quadratic drives preserve global parity.
Terms containing odd powers of individual modes, including intermode pair drives and mirror-pair conversions, generally break the independent modal parities. Whether they directly force a dormant mode depends on the occupied support.
Removing them gives the $(\mathbb Z_2)^N$ symmetry and invariant supports of the effective model.
Small corrections to macroscopic amplitudes can nevertheless exceed the exponentially decayed seeds, and higher-order combinations of off-resonant terms can generate slow forcing.
Residual parity breaking can therefore impose a seed floor, as a coherent drive does in Appendix~\ref{app:numerical_methods}. Preserving the deterministic dwell growth thus requires suppressing these terms below the decayed seeds, a stricter condition than the one set by the mean-field amplitudes.

\subsection{\texorpdfstring{Detuning ladder and drive cutoff}{Detuning ladder and drive cutoff}}
\label{app:ladder_cutoff}
For the ladder of Eq.~\eqref{eq:detuning_ladder}, with $J>0$ and $N\ge2$, adjacent detunings are separated by
\begin{equation}
 \begin{aligned}
 \Delta_m-\Delta_{m+1}
 =4J\sin\!\left(\frac{\pi}{2(N+1)}\right)
 \sin\!\left(\frac{(2m+1)\pi}{2(N+1)}\right).
 \end{aligned}
\label{eq:appA_ladder_gap}
\end{equation}
The vacuum-instability boundaries of two adjacent modes cross where $\Delta_m=-\Delta_{m+1}=\chi$, that is, at $G^2=\kappa^2/4+[(\Delta_m-\Delta_{m+1})/2]^2$.
The largest spacing is the central gap for even $N$, or either gap next to the central mode for odd $N$.
Equation~\eqref{eq:gmax_adjacent_crossing} therefore gives
\begin{equation}
 G_{\max}^{(N)}=
 \begin{cases}
 \displaystyle\sqrt{\frac{\kappa^2}{4}+4J^2\sin^2\!\left(\frac{\pi}{2(N+1)}\right)},
 &N\ \mathrm{even},\\[3mm]
 \displaystyle\sqrt{\frac{\kappa^2}{4}+J^2\sin^2\!\left(\frac{\pi}{N+1}\right)},
 &N\ \mathrm{odd}.
 \end{cases}
\label{eq:appA_ladder_cutoff}
\end{equation}
For fixed $J$ and $\kappa>0$, the leading large-$N$ behavior is the same for both parities,
\begin{equation}
 G_{\max}^{(N)}=\frac\kappa2+\frac{\pi^2J^2}{\kappa(N+1)^2}
 +\mathcal O\!\left((N+1)^{-4}\right).
\label{eq:appA_ladder_cutoff_asymptotic}
\end{equation}
This expansion holds for $N+1\gg2\pi J/\kappa$. The second term is the $N^{-2}$ closing of the drive window used to normalize $g_N$ in Appendix~\ref{app:maps}, where its size dependence is also shown.
The cutoff marks the closure of the last vacuum-stable gap between tongue centers; smaller gaps close at lower drives, and transverse activation on a populated support depends instead on its cross-Kerr-shifted detunings.

\subsection{\texorpdfstring{Occupation scales}{Occupation scales}}
\label{app:occupation_noise}
At fixed detunings, drives, and losses, the mean-field equations preserve their form under the rescaling
\begin{equation}
 \alpha_{m,\rm ph}=\sqrt{n_s}\,\alpha_m,
 \qquad V_{m,n,\rm ph}=\frac{V_{m,n}}{n_s}.
\label{eq:appA_occupation_scale}
\end{equation}
We define the occupation scale $n_s=U/U_{\rm ph}$, the ratio of the reference Kerr scale $U$ to its physical value $U_{\rm ph}$ per excitation. The self-Kerr coefficient of mode $m$ is then $U_{m,\rm ph}=U_m/n_s$, and its mean-field occupation is $n_sI_m$.
Increasing $n_s$ at fixed intensities makes the mean-field description exact for the active modes, whose relative fluctuations scale as $n_s^{-1/2}$. Dormant amplitudes, however, decay far below the active ones, so the growth of successive dwell times additionally requires fluctuations and residual forcing to remain below the inherited seeds (Appendix~\ref{app:rwa_parities}).

\section{Invariant supports, stationary branches, and linear stability}
\label{app:supports}

This appendix derives the invariant subspaces, stationary branches, fluctuation Hamiltonian, Morse-index bound, and transverse and internal spectra used in Secs.~\ref{sec:model}--\ref{sec:landscapes}. We use the mean-field flow of Eq.~\eqref{eq:mf_general_complex} with uniform drive $G$ and loss $\kappa$, and the symmetric Kerr matrix $\mathbf V$ defined there, with $V_{m,m}=U_m$.
The matrix of the many-mode calculations, $\mathbf V=\tfrac{U}{N+1}\boldsymbol\Gamma$, is given in Appendix~\ref{app:model_validity}.

\subsection{\texorpdfstring{Invariant supports}{Invariant supports}}
\label{app:invariant_supports}
For a modal support $S\subseteq\{1,\ldots,N\}$, the associated invariant subspace introduced in Sec.~\ref{sec:chains} is
\begin{equation}
 \mathcal M_S=\left\{\boldsymbol\alpha\in\mathbb C^N:
 \alpha_m=0\ \text{for every }m\notin S\right\}.
\label{eq:appB_support_subspace}
\end{equation}
For a dormant mode $d\notin S$, every term in Eq.~\eqref{eq:mf_general_complex} contains $\alpha_d$ or $\alpha_d^*$, so $\alpha_d=0$ implies $\dot\alpha_d=0$.
The subspace $\mathcal M_S$ is therefore invariant under the full flow, cross-Kerr shifts included.
These subspaces are nested by inclusion, since $S'\subseteq S$ implies $\mathcal M_{S'}\subseteq\mathcal M_S$.
A seeded trajectory keeps small nonzero amplitudes outside the occupied support, so the avalanche words of Sec.~\ref{sec:landscapes} describe visits near these invariant subspaces. A transverse instability ends such a visit by amplifying a dormant seed.

\subsection{\texorpdfstring{Stationary branches and physicality}{Stationary branches and physicality}}
\label{app:stationary_branches}
For a stationary amplitude $\alpha_m=\sqrt{I_m}e^{i\phi_m}\ne0$, dividing Eq.~\eqref{eq:mf_general_complex} by $\alpha_m$ gives $G e^{-2i\phi_m}=\Delta_m^{\mathrm{eff}}+i\kappa/2$.
Taking the modulus requires $G\geq\kappa/2$ and yields, with $\chi=\sqrt{G^2-\kappa^2/4}$,
\begin{equation}
 \begin{aligned}
 \Delta_m^{\mathrm{eff}}&=\sigma_m\chi,
 \qquad \sigma_m\in\{+1,-1\},\\
 e^{-2i\phi_m}&=\frac{\sigma_m\chi+i\kappa/2}{G}.
 \end{aligned}
\label{eq:appB_stationary_phase}
\end{equation}
The phase is fixed modulo $\pi$, giving $2^{|S|}$ independent parity images for each intensity solution.
Let $\mathbf V_S$ denote the principal submatrix on the active modes.
Using $\Delta_m^{\mathrm{eff}}=\Delta_m-\sum_{n\in S}V_{m,n}I_n$, the intensities obey
\begin{equation}
 \begin{aligned}
 \mathbf V_S\boldsymbol I_S&=\boldsymbol X_S,
 \qquad X_m=\Delta_m-\sigma_m\chi,\\
 \boldsymbol I_S&=\mathbf V_S^{-1}\boldsymbol X_S
 \quad\text{if }\det\mathbf V_S\ne0.
 \end{aligned}
\label{eq:appB_support_branch}
\end{equation}
A candidate with support exactly $S$ is physical only if $\chi$ is real and every $I_m>0$ for $m\in S$. A vanishing intensity belongs to a smaller support. For the dimer, Eq.~\eqref{eq:appB_support_branch} reduces to Eq.~\eqref{eq:dimer_stationary_intensities}.
The empty support is the vacuum, $\boldsymbol\alpha=0$, which exists at every drive.

When the nonempty support matrices are invertible, each mode is either dormant or active with one of two branch signs, giving $3^N$ algebraic candidates, a number that grows exponentially with $N$~\cite{borovik2024,ameye2025parametric}.
Including the two parity choices for each active mode gives at most $5^N$ FPs, before imposing physicality and removing coincidences at branch boundaries.
For a singular $\mathbf V_S$, the linear system can be inconsistent or admit a continuous family, which this count of isolated candidates excludes.

\subsection{\texorpdfstring{Linearization and Morse-index bound}{Linearization and Morse-index bound}}
\label{app:linearization_morse}
The Jacobian of Eq.~\eqref{eq:jacobian_general} reads $\mathcal J^{(j)}=-i\mathcal H_{\rm BdG}^{(j)}-\tfrac12\boldsymbol\kappa$, where $\boldsymbol\kappa=\operatorname{diag}(\kappa_1,\ldots,\kappa_N,\kappa_1,\ldots,\kappa_N)$ is the loss matrix and $\mathcal H_{\rm BdG}^{(j)}$ is the Bogoliubov--de Gennes matrix of the quadratic fluctuation Hamiltonian around FP $j$~\cite{blaizot1986quantum},
\begin{equation}
\begin{aligned}
 \hat H_{\rm fl}^{(j)}={}&\sum_{m,n}\bigl(V_{m,n}\alpha_m\alpha_n^{*}-\delta_{m,n}\Delta_m^{\mathrm{eff}}\bigr)\,\delta\hat a_m^\dagger\delta\hat a_n\\
 &+\frac12\sum_{m,n}\bigl(V_{m,n}\alpha_m\alpha_n+\delta_{m,n}G_m\bigr)\,\delta\hat a_m^\dagger\delta\hat a_n^\dagger+{\rm H.c.},
\end{aligned}
\label{eq:fluctuation_hamiltonian}
\end{equation}
where $\alpha_m\equiv\alpha_m^{(j)}$ and $\Delta_m^{\mathrm{eff}}\equiv\Delta_m^{\mathrm{eff},(j)}=\Delta_m-\sum_nV_{m,n}|\alpha_n^{(j)}|^2$ is the effective detuning of Eq.~\eqref{eq:effective_detuning} evaluated at the FP.
The first line is the normal part, in which the interactions shift each mode to $\Delta_m^{\mathrm{eff}}$, and the terms $V_{m,n}\alpha_m\alpha_n^*$ hop fluctuations between modes, with amplitudes set by the active modes.
The second line creates and annihilates fluctuations in pairs. Besides the single-mode squeezing of the drive, the terms $V_{m,n}\alpha_m\alpha_n$ create pairs within each mode, $\propto U_m\alpha_m^2$, and between different active modes.
These intensity-dependent interaction terms vanish at the vacuum and grow with the intensities, so a nonempty support can destabilize directions that are stable at $\boldsymbol\alpha^{(\varnothing)}$.

In components, linearizing Eq.~\eqref{eq:mf_general_complex} around FP $j$ gives the dissipative Bogoliubov--de Gennes Jacobian in the Nambu basis of Eq.~\eqref{eq:jacobian_general}:
\begin{equation}
 \begin{aligned}
 \mathcal J^{(j)}&=-i\begin{pmatrix}
 \mathcal K^{(j)}&\mathcal G^{(j)}\\
 -\mathcal G^{(j)*}&-\mathcal K^{(j)*}
 \end{pmatrix}-\frac\kappa2\mathbb I_{2N},\\
 (\mathcal K^{(j)})_{m,n}&=-\Delta_m^{\mathrm{eff},(j)}\delta_{m,n}
 +V_{m,n}\alpha_m^{(j)}\alpha_n^{(j)*},\\
 (\mathcal G^{(j)})_{m,n}&=G\delta_{m,n}
 +V_{m,n}\alpha_m^{(j)}\alpha_n^{(j)}.
 \end{aligned}
\label{eq:appB_bdg_matrix}
\end{equation}
Here $\Delta_m^{\mathrm{eff},(j)}=\Delta_m-\sum_n V_{m,n}I_n^{(j)}$.
The blocks $\mathcal K^{(j)}$ and $\mathcal G^{(j)}$ contain the normal and pairing terms of the fluctuation Hamiltonian in Eq.~\eqref{eq:fluctuation_hamiltonian}~\cite{gomezleon2023drivendissipative,PhysRevA.98.023841}.
When $\alpha_d^{(j)}=0$, all off-diagonal entries coupling mode $d$ to other modes vanish, so the full Jacobian separates into the active block and one transverse block for each dormant mode.

For the Morse-index bound used in Sec.~\ref{sec:model}, let $H_{\rm cl}$ be the normally ordered Hamiltonian of Eq.~\eqref{eq:general_hamiltonian} with $\hat a_m$ replaced by $\alpha_m=u_m+iv_m$.
In the real quadratures $\boldsymbol r=(u_1,v_1,\ldots,u_N,v_N)^T$, the equivalent Jacobian is
\begin{equation}
 \begin{aligned}
 \mathcal J_{\rm R}^{(j)}&=\frac12\mathbb S\,
 \left.\nabla^2 H_{\rm cl}\right|_{\boldsymbol r^{(j)}}
 -\frac\kappa2\mathbb I_{2N},\\
 \mathbb S&=\bigoplus_{m=1}^N
 \begin{pmatrix}0&1\\-1&0\end{pmatrix}.
 \end{aligned}
\label{eq:appB_morse_jacobian}
\end{equation}
The factor $1/2$ follows from $\{u_m,v_n\}=\delta_{m,n}/2$.
The first term is a Hamiltonian matrix: its eigenvalues occur in pairs $\nu,-\nu$, with conjugate partners for nonreal eigenvalues.
Uniform damping shifts each pair to $\nu-\kappa/2,-\nu-\kappa/2$, of which at most one has positive real part.
Thus the Morse index, counted with algebraic multiplicity, satisfies
\begin{equation}
 \mu^{(j)}=\#\{\lambda\in\operatorname{spec}\mathcal J^{(j)}:
 \operatorname{Re}\lambda>0\}\le N.
\label{eq:appB_morse_bound}
\end{equation}
No FP is therefore a source in the full $2N$-dimensional phase space, and the dimer has $\mu\le2$.
The index counts the unstable directions only. Their modal content, their character, internal or transverse to a support, and their gains can change while the index stays fixed, so regions with equal Morse vectors in Fig.~\ref{fig3}(a) can differ in their escape channels.

\subsection{\texorpdfstring{Transverse stability and branch attachment}{Transverse stability and branch attachment}}
\label{app:transverse_stability}
For a dormant mode $d\notin S$, the stationary intensities of Eq.~\eqref{eq:appB_support_branch} give the effective detuning
\begin{equation}
 \begin{aligned}
 \Delta_d^{\mathrm{eff},(S)}
 &=\Delta_d-\mathbf V_{d,S}\boldsymbol I_S\\
 &=\Delta_d-\mathbf V_{d,S}\mathbf V_S^{-1}\boldsymbol X_S,
 \end{aligned}
\label{eq:appB_dressed_detuning}
\end{equation}
where $\mathbf V_{d,S}=\mathbf V_{S,d}^T$ is the row of couplings $V_{d,n}$, $n\in S$, and the second line assumes $\mathbf V_S$ is invertible.
The transverse Jacobian of mode $d$ is the vacuum block of Eq.~\eqref{eq:single_kpo_origin_jacobian} with $\Delta_d$ replaced by $\Delta_d^{\mathrm{eff},(S)}$, giving the spectrum of Eq.~\eqref{eq:transverse_eigs_general}.
For $G\ge\kappa/2$, the dormant mode is linearly stable when $|\Delta_d^{\mathrm{eff},(S)}|>\chi$ and unstable when $|\Delta_d^{\mathrm{eff},(S)}|<\chi$.
Each unstable dormant mode contributes one positive eigenvalue to the Morse index; at the boundary $\Delta_d^{\mathrm{eff},(S)}=\pm\chi$, the transverse eigenvalues are $0$ and $-\kappa$.

For the dimer's odd branch with $\sigma_o=-1$, $I_o=(\Delta_o+\chi)/U_o$, so
\begin{equation*}
 \Delta_e^{\mathrm{eff}}-\chi
 =\frac{U_o\Delta_e-V_{e,o}\Delta_o-(U_o+V_{e,o})\chi}{U_o}.
\end{equation*}
This difference decreases with $\chi$, and its zero gives Eq.~\eqref{eq:dimer_monomode_activation_threshold}: a physical odd branch enters the even-mode gain window through its upper edge when this threshold is positive. For $U_o=U/2$ and $V_{e,o}=U$, $\Delta_e^{\mathrm{eff}}=3J-\Delta-2\chi$, so the unstable interval is $J-\Delta/3<\chi<3J-\Delta$. It opens only for $\chi_{\{o\}\to\{e,o\}}=J-\Delta/3>0$; at equality, the interval collapses.

To locate the branch on $S\cup\{d\}$, we keep the branch signs on $S$ fixed and add $I_d$ to the linear system of Eq.~\eqref{eq:appB_support_branch}.
Eliminating the modified active intensities $\boldsymbol I'_S$ gives
\begin{equation}
 \begin{aligned}
 I_d&=\frac{\Delta_d^{\mathrm{eff},(S)}-\sigma_d\chi}
 {V_{d,d}-\mathbf V_{d,S}\mathbf V_S^{-1}\mathbf V_{S,d}},\\
 \boldsymbol I'_S&=\boldsymbol I_S-\mathbf V_S^{-1}\mathbf V_{S,d}I_d.
 \end{aligned}
\label{eq:appB_branch_attachment}
\end{equation}
For a physical parent branch and nonzero denominator, $I_d=0$ at $\Delta_d^{\mathrm{eff},(S)}=\sigma_d\chi$: the enlarged branch meets the parent exactly where a transverse eigenvalue vanishes, as at the dimer threshold in Sec.~\ref{sec:chains}.
The enlarged branch is physical only where $I_d>0$ and every component of $\boldsymbol I'_S$ remains positive.
Stability of the enlarged branch requires both its internal spectrum (Appendix~\ref{app:internal_stability}) and all remaining transverse blocks to be stable; the destination of an escaping trajectory follows from the global flow (Sec.~\ref{sec:landscapes}).

\subsection{\texorpdfstring{Internal stability on an active support}{Internal stability on an active support}}
\label{app:internal_stability}
For perturbations confined to $\mathcal M_S$, write $\alpha_m=\sqrt{I_m}e^{i\phi_m}$ for each active mode $m\in S$.
Equation~\eqref{eq:mf_general_complex} gives
\begin{equation}
 \begin{aligned}
 \dot I_m&=I_m[-\kappa-2G\sin(2\phi_m)],\\
 \dot\phi_m&=\Delta_m^{\mathrm{eff}}(\boldsymbol I_S)-G\cos(2\phi_m).
 \end{aligned}
\label{eq:appB_intensity_phase_flow}
\end{equation}
At a physical stationary branch, the phases of Eq.~\eqref{eq:appB_stationary_phase} satisfy $G\sin(2\phi_m)=-\kappa/2$ and $G\cos(2\phi_m)=\sigma_m\chi$.
Linearizing in the deviations $\delta I_m$ and $\delta\phi_m$ therefore yields
\begin{equation}
 \begin{aligned}
 \frac{d}{dt}\begin{pmatrix}\delta\boldsymbol I\\\delta\boldsymbol\phi\end{pmatrix}
 &=\begin{pmatrix}0&-4\chi\,\mathcal I_S\\-\mathbf V_S&-\kappa\mathbb I_{|S|}\end{pmatrix}
 \begin{pmatrix}\delta\boldsymbol I\\\delta\boldsymbol\phi\end{pmatrix},\\
 (\mathcal I_S)_{m,n}&=\sigma_mI_m\,\delta_{m,n},
 \end{aligned}
\label{eq:appB_active_jacobian}
\end{equation}
where the intensities in $\mathcal I_S$ are evaluated at the FP.
The change of variables to intensities and phases is nonsingular for $I_m>0$, so this matrix has the same spectrum as the active block of Eq.~\eqref{eq:appB_bdg_matrix}.

For perturbations proportional to $e^{\lambda t}$, eliminating the phase deviations reduces the spectral problem to
\begin{equation}
 \lambda(\lambda+\kappa)\,\delta\boldsymbol I
 =4\chi\,\mathcal I_S\mathbf V_S\,\delta\boldsymbol I .
\label{eq:appB_active_spectrum}
\end{equation}
Each eigenvalue $\theta_h$ of $4\chi\,\mathcal I_S\mathbf V_S$ thus gives a pair of internal exponents,
\begin{equation}
 \lambda_{h,\pm}=-\frac\kappa2
 \pm\sqrt{\frac{\kappa^2}{4}+\theta_h}.
\label{eq:appB_internal_exponents}
\end{equation}
Internal linear stability requires $\operatorname{Re}\lambda_{h,\pm}<0$ for every $h$.
Stability in the full phase space also requires every dormant block in Eq.~\eqref{eq:transverse_eigs_general} to be stable.
An internal instability changes the motion within $\mathcal M_S$ without activating a dormant mode. Appendix~\ref{app:dimer} derives the resulting Hopf condition for the dimer.

\section{\texorpdfstring{Dimer dynamics: incoming seeds, dwell times, and persistent activity}{Dimer dynamics: incoming seeds, dwell times, and persistent activity}}
\label{app:dimer}
This appendix derives the dimer results of Sec.~\ref{sec:chains}, namely the stationary states, activation threshold, and Morse-resolved landscape of Figs.~\ref{fig2} and~\ref{fig3}(a), the incoming seed of Eq.~\eqref{eq:dimer_seed_size} and its oscillating prefactor [Fig.~\ref{fig4}(b)], the plateau dwell of Eq.~\eqref{eq:dimer_inherited_seed_summary} [Fig.~\ref{fig4}(c)], and the Hopf and global bifurcations of the limit cycles along the cut $\Delta/U=3.5$.

\subsection{\texorpdfstring{Dimer model, stationary states, and Morse-resolved landscape}{Dimer model, stationary states, and Morse-resolved landscape}}
\label{app:dimer_morse}
For the dimer of Sec.~\ref{sec:chains}, Eq.~\eqref{eq:general_hamiltonian} reads
\begin{equation}
    \hat H=\sum_{m=e,o}\left[-\Delta_m\hat n_m+\frac{U}{4}\hat a_m^{\dagger2}\hat a_m^2+\frac{G}{2}\left(\hat a_m^{\dagger2}+\hat a_m^2\right)\right]+U\hat n_e\hat n_o.
    \label{eq:dimer_hamiltonian}
\end{equation}
The detuning split thus separates the two gain windows, $|\Delta_e|<\chi$ and $|\Delta_o|<\chi$ [Eq.~\eqref{eq:single_kpo_threshold}].
Where the windows do not overlap, the vacuum is unstable along one mode only, so the modes activate one at a time, the first directly from the vacuum and the second only once the first has shifted it into its window [Fig.~\ref{fig2}(a)].

A dimer has four possible supports, or activity patterns, namely the vacuum $\varnothing$, the monomodes $\{e\}$ and $\{o\}$, and the two-mode support $\{e,o\}$, according to whether no, one, or both modes are active.
At an FP, each active mode $m$ has phase $\phi_{m,\sigma_m}$ as in Eq.~\eqref{eq:single_kpo_stationary_solutions}, with $\sigma_m=\pm1$ labeling its response branch. Its effective detuning includes both self- and cross-Kerr shifts and obeys $\Delta_m^{\mathrm{eff}}=\Delta_m-\sum_aV_{m,a}I_a=\sigma_m\chi_m$ [Eq.~\eqref{eq:effective_detuning}]. Every active mode thus has its effective detuning at the edge of its gain window.
Solving for each support gives
\begin{align}
 \varnothing &: I_e=I_o=0,\nonumber\\
 \{e\} &: I_e=(\Delta_e-\sigma_e\chi_e)/U_e,\quad I_o=0,\nonumber\\
 \{o\} &: I_e=0,\quad I_o=(\Delta_o-\sigma_o\chi_o)/U_o,\nonumber\\
 \{e,o\} &:
 \begin{pmatrix}I_e\\ I_o\end{pmatrix}
 =
 \begin{pmatrix}U_e&V_{e,o}\\V_{o,e}&U_o\end{pmatrix}^{-1}
 \begin{pmatrix}\Delta_e-\sigma_e\chi_e\\\Delta_o-\sigma_o\chi_o\end{pmatrix}.
 \label{eq:dimer_stationary_intensities}
\end{align}
For a monomode, Eq.~\eqref{eq:dimer_stationary_intensities} reduces to Eq.~\eqref{eq:single_kpo_stationary_solutions}; with both modes active, the cross-Kerr shifts couple the two intensities [Appendix~\ref{app:supports}].
The phases remain those of Eq.~\eqref{eq:single_kpo_stationary_solutions}, because the cross-Kerr interaction only shifts the detuning.
A candidate is a stationary state only when it is physical, i.e., when $\chi_m$ is real and every active intensity is positive.

Applying these physicality conditions to the candidates of Eq.~\eqref{eq:dimer_stationary_intensities} across the $(\Delta/U,G/U)$ plane yields the phase diagram of Fig.~\ref{fig2}(b), labeled by the counts $(N_{\varnothing},N_{\{e\},\{o\}},N_{\{e,o\}})$ of physical empty-, one-, and two-mode solutions, stable or not.
Outside both instability tongues, where $|\Delta_e|>\chi$ and $|\Delta_o|>\chi$ [white dashed curves in Fig.~\ref{fig2}(b)], the vacuum is stable but can coexist with stationary branches of nonzero amplitude; entering a tongue destabilizes the vacuum along the corresponding mode, whose monomode branch starts at the tongue boundary.
Even- and odd-mode branches coexist wherever both are physical, even beyond the overlap of the tongues, and the two-mode branches, inside the colored dotted curves of Fig.~\ref{fig2}(b), are analogous to the mixed-mode states of coupled KPOs~\cite{ameye2025parametric,heugel2021ising,heugel2024proliferation}.
Along the exemplary cut $\Delta/U=3.5$, used in Sec.~\ref{sec:chains}, the same expressions give the stationary branches of Fig.~\ref{fig2}(c).

To label an activation threshold by the support change it produces, we write $G^{\rm th}_{\{o\}\to\{e,o\}}$ for the drive at which the even mode activates on the odd branch.
For equal drives and losses, we substitute the odd-branch intensity from Eq.~\eqref{eq:dimer_stationary_intensities}, with $\sigma_o=-1$, into the transverse boundary $\Delta_e-V_{e,o}I_o=\chi$ and obtain
\begin{equation}
\begin{aligned}
    G_{\{o\}\to\{e,o\}}^{\rm th}
    &=
    \sqrt{\chi_{\{o\}\to\{e,o\}}^2+\frac{\kappa^2}{4}}\,,\\
    \chi_{\{o\}\to\{e,o\}}
    &=
    \frac{U_o\Delta_e-V_{e,o}\Delta_o}{U_o+V_{e,o}}\,,
\end{aligned}
    \label{eq:dimer_monomode_activation_threshold}
\end{equation}
provided that the parent odd branch is physical and $\chi_{\{o\}\to\{e,o\}}>0$, i.e., that the odd intensity pulls $\Delta_e^{\rm eff}$ down into the gain window from above (Appendix~\ref{app:supports}). At $\Delta/U=3.5$, $G^{\rm th}_{\{o\}\to\{e,o\}}\simeq3.10\,U$.

A dimer FP can be unstable along at most two directions (Appendix~\ref{app:supports}), so we collect the FP counts in the three-component Morse vector $\boldsymbol n=(n_0,n_1,n_2)$, with $n_\mu$ the number of FPs with $\mu$ unstable directions: $n_0$ counts stable FPs and $n_1,n_2$ saddles.
Resolving the counts by Morse index distinguishes regions with the same stationary support counts [Figs.~\ref{fig2}(b) and~\ref{fig3}(a)].
Since the Morse vectors do not resolve which directions are unstable, internal or transverse, or how fast they grow (Appendix~\ref{app:supports}), regions with identical counts can still differ in their escape channels and rates, and these differences select the subsequent support changes.
Along the cut, the Morse vectors of Fig.~\ref{fig3}(a) reflect the outcomes of Fig.~\ref{fig3}(c)--(e). At $G/U=3.26$, $\boldsymbol n=(6,5,0)$. The four mixed FPs and the two even phase states are stable, while the vacuum, the odd phase states, and the $\sigma=+1$ even states are saddles. At $G/U=3.49$ and $4.25$, the mixed FPs have become saddles with two unstable directions, giving $\boldsymbol n=(2,5,4)$ and leaving only the even phase states stable.
These two drives thus share the same Morse vector but realize different outcomes, the limit cycle and the even monomode, so the Morse vector alone does not fix the avalanche word.

\subsection{\texorpdfstring{Initial perturbation and local coordinates}{Initial perturbation and local coordinates}}
\label{app:dimer_coordinates}
We use the effective dimer of Eq.~\eqref{eq:dimer_hamiltonian}, with modes $e,o$, detunings $\Delta_e=\Delta+J$ and $\Delta_o=\Delta-J$, and common drive $G$ and loss $\kappa$.
Writing $\alpha_m=u_m+iv_m$ and $\boldsymbol r=(u_e,v_e,u_o,v_o)^T$, the initial perturbation is
\begin{equation}
 \boldsymbol r(0)=\varepsilon_0\hat{\boldsymbol e},
 \qquad \|\hat{\boldsymbol e}\|_2=1,\quad 0<\varepsilon_0\ll1.
\label{eq:appC_dimer_initial_seed}
\end{equation}
At the vacuum and the odd monomode, the Jacobian separates into even and odd blocks.
For either support $S$, with FP $\boldsymbol r_S$ in real quadratures, let $\boldsymbol v_\pm^m(S)$ and $[\boldsymbol w_\pm^m(S)]^T$ be the right eigenvectors and dual left rows of the Jacobian at $S$, with eigenvalues $\lambda_\pm^m(S)$ and $m=e,o$.
Away from eigenvector coalescences, we choose
\begin{equation}
 \begin{aligned}
 [\boldsymbol w_a^m(S)]^T\boldsymbol v_b^n(S)
 &=\delta_{mn}\delta_{ab},\qquad a,b=\pm,\\
 c_a^m(S,t)&=[\boldsymbol w_a^m(S)]^T
 [\boldsymbol r(t)-\boldsymbol r_S].
 \end{aligned}
\label{eq:appC_dimer_biorthogonal_coordinate}
\end{equation}
We normalize the growing right eigenvectors to unit Euclidean norm in real quadratures and define their duals and the exit amplitude $c_\star$ in the same convention.
Transforming the right vectors and their duals together into Nambu components leaves the scalar coordinate unchanged, giving the projection used in Sec.~\ref{sec:chains}.

The incoming seed for the transverse escape is
\begin{equation}
 c_{+,e}^{\mathrm{in}}
 =c_+^e(\{o\},t_{\mathrm{in}}),
\label{eq:appC_dimer_incoming_coordinate}
\end{equation}
where $t_{\mathrm{in}}$ is the arrival time at the saddle's neighborhood.
Because the support change from the vacuum to the odd plateau takes a finite time, $t_{\mathrm{in}}$ exceeds $\tau(\varnothing)$. The transfer row of Appendix~\ref{app:dimer_seed} accounts for this support change.

\subsection{\texorpdfstring{Vacuum decay and incoming seed}{Vacuum decay and incoming seed}}
\label{app:dimer_seed}
In the vacuum linearization, the growing odd coordinate reaches its fixed exit amplitude $c_\star$ after
\begin{equation}
 \tau(\varnothing)\simeq\frac{1}{\lambda_+^o(\varnothing)}
 \ln\!\left[\frac{c_\star}{|c_+^o(\varnothing,0)|}\right].
\label{eq:appC_dimer_vacuum_dwell}
\end{equation}
For a fixed initial direction with nonzero projection onto the growing odd eigenvector, $|c_+^o(\varnothing,0)|\propto\varepsilon_0$.
The vacuum dwell therefore grows as $\ln(1/\varepsilon_0)$, as in Eq.~\eqref{eq:ring_up_time}.

Let $\boldsymbol r_e(0)=\varepsilon_0\hat{\boldsymbol e}_e$ contain the initial even quadratures.
Their vacuum block and its decay rate are
\begin{equation}
 \begin{aligned}
 \mathcal J_{{\rm R},e}^{(\varnothing)}&=\begin{pmatrix}
 -\kappa/2&-(\Delta_e+G)\\
 \Delta_e-G&-\kappa/2
 \end{pmatrix},\\
 \gamma^e(\varnothing)&=-\max_\pm\operatorname{Re}
 \lambda_\pm^e(\varnothing)>0.
 \end{aligned}
\label{eq:appC_dimer_even_contraction}
\end{equation}
We define the row $\boldsymbol{\mathcal T}_0^e$, which propagates a small even displacement along the reference support change from the end of the vacuum dwell to the odd-saddle neighborhood and projects it onto the dual of the outgoing even eigenvector.
At fixed entry and exit neighborhoods, we treat this reference transfer as independent of $\varepsilon_0$ to leading order.
Keeping the full stable vacuum propagator gives
\begin{equation}
 \begin{aligned}
 c_{+,e}^{\mathrm{in}}
 &\simeq\boldsymbol{\mathcal T}_0^e
 e^{\mathcal J_{{\rm R},e}^{(\varnothing)}\tau(\varnothing)}
 \boldsymbol r_e(0)\\
 &=\varepsilon_0e^{-\gamma^e(\varnothing)\tau(\varnothing)}
 f_e(\tau(\varnothing);\hat{\boldsymbol e}).
 \end{aligned}
\label{eq:appC_dimer_inherited_seed}
\end{equation}
For all three drives of Fig.~\ref{fig4}, $\Delta_e^2>G^2$, so
$\gamma^e(\varnothing)=\kappa/2$ and
$\nu_e=\sqrt{\Delta_e^2-G^2}$.
The propagator is a damped rotation in a nonorthogonal basis, yielding
\begin{equation}
 f_e(t;\hat{\boldsymbol e})=a_e\cos(\nu_e t)+b_e\sin(\nu_e t),
\label{eq:appC_dimer_projection_factor}
\end{equation}
where $a_e,b_e$ depend on the initial direction and the reference transfer.
Substituting Eq.~\eqref{eq:appC_dimer_vacuum_dwell} gives
\begin{equation}
 |c_{+,e}^{\mathrm{in}}|\propto
 \varepsilon_0^{\,1+\gamma^e(\varnothing)/\lambda_+^o(\varnothing)}
 |f_e(\tau(\varnothing);\hat{\boldsymbol e})|.
\label{eq:appC_dimer_seed_power}
\end{equation}
The rates fix the power-law envelope of Eq.~\eqref{eq:dimer_seed_size}, while the projection oscillates with $\ln\varepsilon_0$ [Fig.~\ref{fig4}(b)].
This leading-order estimate holds away from zeros of the projected linear seed, where higher-order contributions take over.

\subsection{\texorpdfstring{History-dependent dwell on the odd plateau}{History-dependent dwell on the odd plateau}}
\label{app:dimer_dwell}
Near the odd saddle, the incoming coordinate grows with the transverse gain $\lambda_+^e(\{o\})>0$.
The local first-passage law is Eq.~\eqref{eq:dimer_inherited_seed_summary}, with the dwell timed from entry into the saddle's neighborhood and the fixed exit amplitude $c_\star$ of Sec.~\ref{sec:chains}.
Figure~\ref{fig4}(c) expresses the dwell in growth decades by dividing $\lambda_+^e\tau$ by $\ln10$.
Using Eq.~\eqref{eq:appC_dimer_inherited_seed},
\begin{equation}
 \begin{aligned}
 \tau(\{o\})\simeq\frac{1}{\lambda_+^e(\{o\})}
 \bigg\{&\ln\!\left[\frac{c_\star}
 {\varepsilon_0|f_e(\tau(\varnothing);\hat{\boldsymbol e})|}\right]\\
 &+\gamma^e(\varnothing)\tau(\varnothing)\bigg\}.
 \end{aligned}
\label{eq:appC_dimer_inherited_dwell}
\end{equation}
Equation~\eqref{eq:appC_dimer_inherited_dwell} is the $k=1$ case of Eq.~\eqref{eq:general_inherited_dwell_law}. The preceding vacuum dwell enters through the decay term $\gamma^e(\varnothing)\tau(\varnothing)$, and the rotation accumulated during that dwell enters through $\ln|f_e|$.
The local law, Eq.~\eqref{eq:dimer_inherited_seed_summary}, instead uses the actual incoming projection and is the comparison made in Fig.~\ref{fig4}(c).

\Needspace{8\baselineskip}
\subsection{\texorpdfstring{Internal Hopf instability of the mixed support}{Internal Hopf instability of the mixed support}}
\label{app:dimer_hopf}
The internal spectral problem is Eq.~\eqref{eq:appB_active_spectrum} of Appendix~\ref{app:supports}.
For the physical mixed support, set $\Lambda=\operatorname{diag}(\sqrt{I_e},\sqrt{I_o})$.
If the branch signs agree, $\sigma_e=\sigma_o=s$, then
\begin{equation}
 \Lambda^{-1}(4\chi\,\mathcal I_S\mathbf V_S)\Lambda
 =4s\chi\,\Lambda\mathbf V_S\Lambda.
\label{eq:appC_dimer_equal_signs}
\end{equation}
This matrix is real symmetric, so its eigenvalues $\theta$ are real.
Equation~\eqref{eq:appB_internal_exponents} then gives real exponents or complex pairs with real part $-\kappa/2$; no finite-frequency pair can cross the imaginary axis.
Equal-sign branches can therefore lose internal stability only through a zero eigenvalue.

For opposite signs, $\sigma_e=s=-\sigma_o$, the transformed matrix is instead $4\chi\operatorname{diag}(s,-s)\Lambda\mathbf V_S\Lambda$, which need not have real eigenvalues.
Using the dimer coefficients of Eq.~\eqref{eq:dimer_hamiltonian}, the stationary intensities and the associated discriminant are
\begin{equation}
 \begin{aligned}
 I_e^{(s)}&=\frac{2(\Delta-3J+3s\chi)}{3U},\\
 I_o^{(s)}&=\frac{2(\Delta+3J-3s\chi)}{3U},\\
 \Xi_{e,o}^{(s)}
 &=4U^2I_e^{(s)}I_o^{(s)}
 -\frac{U^2}{4}(I_e^{(s)}+I_o^{(s)})^2\\
 &=\frac43\Delta^2-16(J-s\chi)^2.
 \end{aligned}
\label{eq:appC_dimer_hopf_q}
\end{equation}
When $\Xi_{e,o}^{(s)}>0$, the two eigenvalues are
\begin{equation}
 \theta_\pm=4\chi(\chi-sJ)\pm2i\chi\sqrt{\Xi_{e,o}^{(s)}}.
\label{eq:appC_dimer_hopf_eigenvalues}
\end{equation}
At an oscillatory crossing, $\lambda=i\omega_H$ implies $\theta=-\omega_H^2+i\kappa\omega_H$.
Equating real and imaginary parts yields
\begin{equation}
 \begin{aligned}
 \frac43\Delta^2-16(J-s\chi)^2
 &=\frac{\kappa^2}{\chi}(sJ-\chi),\\
 \omega_H&=2\sqrt{\chi(sJ-\chi)}.
 \end{aligned}
\label{eq:appC_dimer_hopf_boundary}
\end{equation}
The branch must satisfy $I_e^{(s)},I_o^{(s)}>0$, $\chi>0$, $\Xi_{e,o}^{(s)}>0$, and $sJ>\chi$.
For $(\Delta,J,\kappa)/U=(3.5,3,5)$, the $s=+1$ branch crosses at $G_H/U\simeq3.38$.
The local spectral condition fixes the onset and the frequency $\omega_H$. Beyond the Hopf point, we identify the stable cycles by long-time integration and periodic-orbit continuation (Appendix~\ref{app:numerical_methods}). These cycles are the persistent two-mode activity of Fig.~\ref{fig3}(d).

\subsection{\texorpdfstring{Symmetries and global bifurcations of the cycles}{Symmetries and global bifurcations of the cycles}}
\label{app:dimer_cycles}
\begin{figure}[!t]
\centering
\includegraphics[width=\columnwidth]{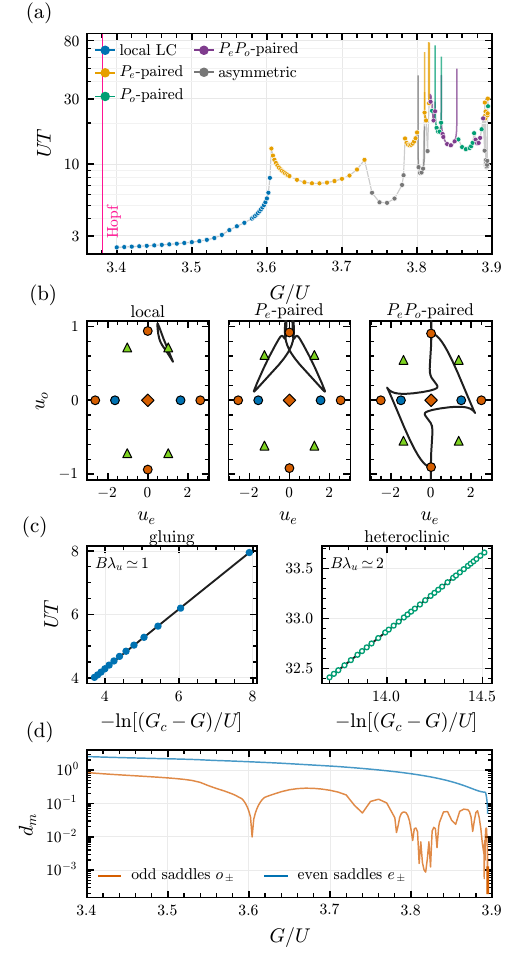}
\caption{Global organization of the dimer cycles at $(\Delta,J,\kappa)/U=(3.5,3,5)$.
(a)~Period versus drive, colored by half-period symmetry; points and thin gray guide, long-time integration; branch curves, continuation.
(b)~Local, $P_e$-paired, and $P_eP_o$-paired examples [parities and pairing of Eqs.~\eqref{eq:appC_dimer_parities} and~\eqref{eq:appC_dimer_half_period}] at $G/U=3.50,3.70,3.84$, respectively; FP colors and marker shapes as in Fig.~\ref{fig:1}(d).
(c)~Logarithmic period fits near the pairwise gluing and the end of the oscillatory regime, with $UT=A-B\ln[(G_c-G)/U]$ and dimensionless slope $B\lambda_u$, where $\lambda_u=\lambda_+^e(\{o\})/U$ at the respective critical drive.
(d)~Minimum Euclidean distances $d_m$, $m=e,o$, in quadratures from each orbit to the even and odd monomode saddle families, minimized over time and family members; $e_\pm,o_\pm$ label the two parity images on each stationary branch.}
\label{fig:appC_global_organization}
\end{figure}
Beyond the Hopf point, the cycles undergo two global bifurcations, a pairwise gluing at $G/U\simeq3.60$ and the end of the oscillatory regime at $G/U\simeq3.89$, where the last family approaches a heteroclinic cycle between the odd-monomode saddles.
The dimer parities act on the quadratures as
\begin{equation}
 \begin{aligned}
 P_e(u_e,v_e,u_o,v_o)&=(-u_e,-v_e,u_o,v_o),\\
 P_o(u_e,v_e,u_o,v_o)&=(u_e,v_e,-u_o,-v_o).
 \end{aligned}
\label{eq:appC_dimer_parities}
\end{equation}
A parity can map a periodic orbit to a distinct symmetry partner or act within the same orbit as a time shift.

For an orbit of period $T$, we test the half-period relation
\begin{equation}
 \boldsymbol r(t+T/2)=g\boldsymbol r(t),\qquad
 g\in\{P_e,P_o,P_eP_o\}.
\label{eq:appC_dimer_half_period}
\end{equation}
When this holds, the parity acts within the orbit as a half-period shift.
We call a cycle $g$-paired when it obeys Eq.~\eqref{eq:appC_dimer_half_period}, local when it remains in a single parity sector, and asymmetric when it visits several sectors without any half-period symmetry [Fig.~\ref{fig:appC_global_organization}(a),(b)].
Long-time integration and continuation identify the same families on overlapping drive intervals (Appendix~\ref{app:numerical_methods}).

We diagnose the global transitions by the period and by the distance to the saddle families.
A slow passage near a saddle gives the logarithmic form
\begin{equation}
 UT=A-B\ln[(G_c-G)/U],
\label{eq:appC_dimer_period_divergence}
\end{equation}
where $A$ and $B$ are the fitted offset and logarithmic slope, and $G_c$ is the critical drive of the event being fitted.
Period growth accompanied by approach to a saddle distinguishes these global transitions from a finite-period local instability.
Such reconnections reorganize the parity sectors visited by the flow, as in the self-oscillatory systems of Ref.~\cite{gomez2025quantum}.

At $G_1/U\simeq3.60$, the four local cycles glue in pairs through the odd-monomode saddles, giving two $P_e$-paired cycles.
The fitted slope $B\lambda_+^e(\{o\})/U\simeq1.02$ is consistent with one slow passage per primitive period on approach to this event [Fig.~\ref{fig:appC_global_organization}(c)].
At $G_c/U\simeq3.89$, the terminal $P_o$-paired family approaches a heteroclinic cycle connecting the two parity images of the odd saddle on the same stationary branch. The 35-point fit over $5.00\times10^{-7}\le(G_c-G)/U\le1.12\times10^{-6}$ gives $B\lambda_+^e(\{o\})/U=2.112\pm0.003$ (formal 95\% fit interval); using the last 20--50 points gives $2.09$--$2.14$, indicating finite-window sensitivity.
The available checks show a much closer approach to the odd saddles, without establishing the limiting distance to the even family [Fig.~\ref{fig:appC_global_organization}(d)].

Between these events, $P_e$-, $P_o$-, $P_eP_o$-paired, and asymmetric cycles occur.
In particular, the four-sector example at $G/U=3.84$ is $P_eP_o$-paired, distinct from the terminal $P_o$ family.
Together, the continuation, period, and saddle-distance diagnostics establish the first pairwise gluing and the terminal heteroclinic cycle. The connections between these two events, including the origin of the four-sector cycle, remain unresolved, because long-time integration can jump between coexisting branches.

\FloatBarrier

\section{\texorpdfstring{Outcome maps of many-mode networks}{Outcome maps of many-mode networks}}
\label{app:maps}
This appendix maps the finite-time outcomes of the ladder networks of Sec.~\ref{sec:landscapes}, from which the monomode regime of the aging avalanche is chosen.

\begin{figure*}[t]
\centering
  \includegraphics[width=\textwidth]{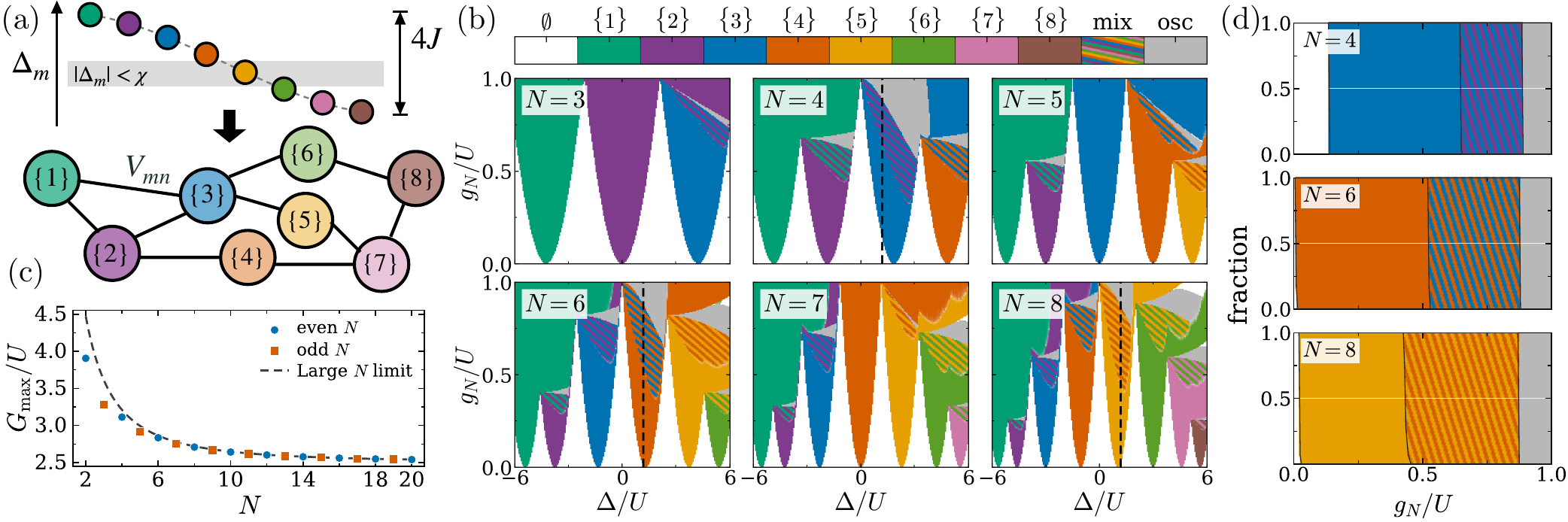}
  \caption{(a)~Top, detuning ladder of Eq.~\eqref{eq:detuning_ladder} for $N=8$, inside the interval $\Delta\pm2J$ of width $4J$, with the gain window $|\Delta_m|<\chi$ shaded; bottom, modal network; nodes, single-mode supports $\{m\}$, colored as in (b); links, selected modal couplings $V_{m,n}$ (all pairs are coupled).
(b)~Dominant finite-time outcomes for $N=3,\ldots,8$ in $(\Delta/U,g_N/U)$, using 32 uniformly sampled initial directions per point; white, endpoints classified as vacuum; solid colors, single-mode supports; stripes, mixed supports; gray, persistent activity. Unresolved trajectories are colored by their endpoint support. Dashed lines: $\Delta/U=1.20$ cuts for $N=4,6,8$.
(c)~Size-dependent cutoff $G_{\max}^{(N)}$, defined by the highest intersection of two adjacent linear vacuum-instability boundaries [Eq.~\eqref{eq:gmax_adjacent_crossing}].
(d)~Outcome fractions across the same 32-member ensembles along the cuts of (b).
Parameters: $\kappa/U=5$, $J/U=3$, $\varepsilon_0=10^{-3}$, integration horizon $UT=5000$, and time step $U\,dt=0.02$.}
  \label{fig:many_mode_landscapes}
\end{figure*}

Each mode can activate from the vacuum only for drives inside its instability tongue, $|\Delta_m|<\chi$ [Eq.~\eqref{eq:single_kpo_threshold}]. Where the tongues do not overlap, at most one mode is unstable at the vacuum.
Adjacent tongues, centered at $\Delta_m=0$ and $\Delta_{m+1}=0$, overlap once the drive exceeds the crossing of their boundaries; the highest such crossing,
\begin{equation}
G_{\max}^{(N)}=
\sqrt{\left(\frac{\kappa}{2}\right)^2+
\left[\frac{1}{2}\max_{1\leq m<N}\left(\Delta_m-\Delta_{m+1}\right)\right]^2},
\label{eq:gmax_adjacent_crossing}
\end{equation}
where the spacing $\Delta_m-\Delta_{m+1}$ is independent of $\Delta$, marks where the last vacuum-stable gap between adjacent tongue centers closes.
As $N$ grows, more modes fit into the same bandwidth $4J$, adjacent tongues move closer, and the cutoff decreases toward the loss threshold $\kappa/2$, so the drive window below it closes as $1/N^2$ [Fig.~\ref{fig:many_mode_landscapes}(c); Appendix~\ref{app:model_validity}].
We therefore express the drive through $g_N\equiv U\,(G-\kappa/2)/(G_{\max}^{(N)}-\kappa/2)$, so that $g_N/U$ runs from the loss threshold ($g_N/U=0$) to the cutoff ($g_N/U=1$) for every $N$.

We now follow the dynamics across this landscape and record the finite-time outcomes.
Figure~\ref{fig:many_mode_landscapes}(b) extends the dimer outcome map of Fig.~\ref{fig3}(b) to $N=3,\ldots,8$. Each panel records the state reached from the vacuum after a fixed integration time.
At small $g_N/U$, the map splits into the $N$ vacuum-instability tongues, centered at $\Delta=-2J\cos k_m$; inside each, the trajectory reaches that mode's monomode by the end of the run, as in the dimer.
Raising $g_N/U$ shifts neighboring modes into their gain windows, as above $G^{\rm th}_{\{o\}\to\{e,o\}}$ in the dimer, and monomode regions give way to mixed supports (stripes) and persistent activity (gray), the many-mode counterpart of the dimer limit cycle of Fig.~\ref{fig3}(d).
Between any two adjacent tongues the dimer sequence recurs (monomode, mixed, persistent activity), so under this normalization the panels for $N=3$--$8$ show qualitatively similar patterns, with outcomes that involve mainly two neighboring modes.

Not every endpoint represents a stationary attractor: the white band at the upper right of Fig.~\ref{fig:many_mode_landscapes}(b) for $N=6$--$8$ lies inside the instability tongue of the highest mode, where the vacuum is a saddle. These near-vacuum endpoints are consistent with cyclic avalanches passing close to the vacuum; Sec.~\ref{sec:many_mode_aging} shows an example of such a heteroclinic cycle (Appendix~\ref{app:deterministic_methods}).
Along the cut $\Delta/U=1.20$, increasing $g_N/U$ takes $N=4,6,8$ from a dominant single-mode outcome through mixed supports to persistent activity [Fig.~\ref{fig:many_mode_landscapes}(d)]. With $N$, the monomode range shrinks and the mixed range widens, whereas persistent activity first appears at nearly the same drive, $g_N/U=0.890$, $0.880$, and $0.875$ for $N=4$, $6$, and $8$.
A trajectory can visit many monomode supports before settling, as we analyze in Sec.~\ref{sec:many_mode_aging}.
Internal instabilities, which involve the hopping and squeezing channels shared by the active modes~\cite{gomezleon2023drivendissipative,PhysRevA.98.023841}, can instead produce persistent activity without changing the support, as in Sec.~\ref{sec:chains} (Appendix~\ref{app:supports}).
For the $N=20$ avalanche of Sec.~\ref{sec:many_mode_aging}, we take $J/U=1$ and $\kappa/U=0.057$. Its drive, $g_{20}/U\simeq0.36$, lies in the monomode regime below the cutoff, outside the parameter range of Fig.~\ref{fig:many_mode_landscapes}.

\section{\texorpdfstring{Accumulated decay, dwell growth, and cyclic avalanches}{Accumulated decay, dwell growth, and cyclic avalanches}}
\label{app:deterministic_methods}
This appendix supports Sec.~\ref{sec:many_mode_aging}. It derives the dwell law of Eq.~\eqref{eq:general_inherited_dwell_law}, the delay product of Eq.~\eqref{eq:spectral_delay_product} and its comparison with the computed dwells [Fig.~\ref{fig:appE_dwell_laws}], and the logarithmic event count, and it identifies the saddle connections of the cyclic avalanche of Fig.~\ref{fig:complex_cascades}(b).

\subsection{\texorpdfstring{Incoming seeds and accumulated decay}{Incoming seeds and accumulated decay}}
\label{app:accumulated_contraction}
We consider the successive supports $S_0=\varnothing,S_1,\ldots,S_k$, with $d_k$ the mode that triggers escape from $S_k$ and $\lambda_k=\lambda_+^{d_k}(S_k)>0$, written $\lambda_+(S_k)$ in Eq.~\eqref{eq:spectral_delay_product}.
We use the incoming projections and eigenvector normalization of Appendix~\ref{app:dimer}, with the exit amplitude $c_\star$ in the same coordinate.
The seed inherited at the arrival time $t_{\mathrm{in}}(S_k)$ is
\begin{equation}
 |c_{+,d_k}^{\mathrm{in}}(S_k)|\simeq
 \varepsilon_0|\mathcal T_k^{d_k}|
 e^{-\int_0^{t_{\mathrm{in}}(S_k)}\gamma^{d_k}(t)\,dt},
\label{eq:appD_inherited_seed}
\end{equation}
where the dimensionless factor $\mathcal T_k^{d_k}$ includes the fixed initial direction, rotations, projections, and transfer between neighborhoods.
At an earlier FP, $|\Delta_{d_k}^{\mathrm{eff}}|>G$ gives a complex transverse pair with real part $-\kappa/2$ [Eq.~\eqref{eq:transverse_eigs_general}], hence a decay envelope with $\gamma^{d_k}=\kappa/2$.
We extend this rate through the support changes and absorb deviations from it into the transfer factor.
For the avalanche of Fig.~\ref{fig:many_mode_aging}, the condition $|\Delta_{d_k}^{\mathrm{eff}}|>G$ holds on every earlier support, although only by 6--14\% for the next two escape modes. Near this boundary, non-normal transient growth enters the transfer factors.

\begin{figure}[!htb]
\centering
\includegraphics[width=\columnwidth]{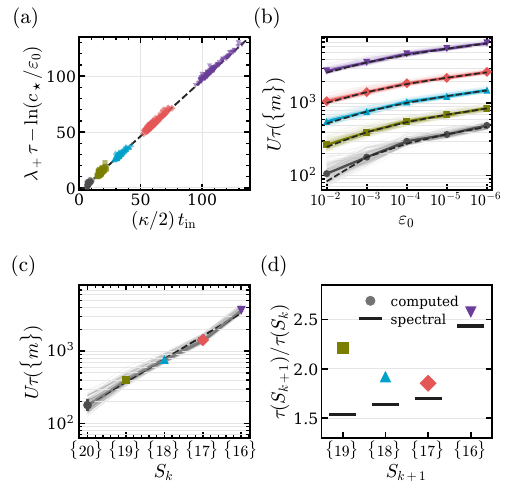}
\caption{Dwell laws of the $N=20$ avalanche of Fig.~\ref{fig:many_mode_aging}.
(a)~Arrival-time law for all 640 escapes at $\varepsilon_0=10^{-3}$: $\lambda_k\tau(S_k)-\ln(c_\star/\varepsilon_0)$ versus $(\kappa/2)t_{\mathrm{in}}(S_k)$; dashed line, unit slope and median offset $-3.73$.
(b)~Dwells versus $\varepsilon_0$ for the 128 initial directions; thin curves, individual dwells; bands, 16th--84th percentiles; symbols, medians; dashed lines, slopes from Eq.~\eqref{eq:spectral_delay_product}, anchored to each plateau's median at $\varepsilon_0=10^{-3}$.
(c)~Five-dwell sequences of individual trajectories (thin gray lines), medians (symbols), 16th--84th percentiles (bars), and geometric fit with $q=2.075$ (dashed).
(d)~Ratios of consecutive median dwells (computed, markers colored by the later plateau $S_{k+1}$) and spectral predictions for ratios of delays per decade $(\lambda_k+\kappa/2)/\lambda_{k+1}$ (black bars).
Panels follow Appendices~\ref{app:accumulated_contraction}--\ref{app:log_event_spacing}: (a) the accumulated decay, (b) the delays per decade, (c) the dwell sequence, and (d) its ratios. Plateau colors and markers as in Fig.~\ref{fig:many_mode_aging}.}
\label{fig:appE_dwell_laws}
\end{figure}

Amplification of the incoming seed to $c_\star$ gives
\begin{equation}
 \tau(S_k)\simeq\frac{1}{\lambda_k}
 \left[\ln\frac{c_\star}{\varepsilon_0|\mathcal T_k^{d_k}|}
 +\int_0^{t_{\mathrm{in}}(S_k)}\gamma^{d_k}(t)\,dt\right].
\label{eq:appD_inherited_dwell}
\end{equation}
Equation~\eqref{eq:appD_inherited_dwell} is the dwell law of Eq.~\eqref{eq:general_inherited_dwell_law}, with the logarithmic correction $-\ln|\mathcal T_k^{d_k}|$ made explicit. The local escape remains exponential, and its starting seed retains the preceding decay.
With the uniform rate $\kappa/2$, Eq.~\eqref{eq:appD_inherited_dwell} gives the arrival-time law $\lambda_k\tau(S_k)-\ln(c_\star/\varepsilon_0)\simeq(\kappa/2)t_{\mathrm{in}}(S_k)$ up to the transfer term. All 640 escapes follow it with a common finite offset [Fig.~\ref{fig:appE_dwell_laws}(a)].

\subsection{\texorpdfstring{Spectral prediction of the delays per decade}{Spectral prediction of the delays per decade}}
\label{app:spectral_delays}
We call $L_k$ the delay coefficient, the slope of $\tau(S_k)$ in $\ln(1/\varepsilon_0)$ that enters Eq.~\eqref{eq:spectral_delay_product}.
If the support changes and transfer factors contribute no leading term in $\ln(1/\varepsilon_0)$, Eq.~\eqref{eq:appD_inherited_dwell} gives
\begin{equation}
 L_0=\frac{1}{\lambda_0},\qquad
 L_k=\frac{1}{\lambda_k}
 \left[1+\sum_{j<k}\gamma^{d_k}(S_j)L_j\right],
\label{eq:appD_delay_recursion}
\end{equation}
where $\lambda_0=\lambda_+^{20}(\varnothing)$ for the avalanche of Fig.~\ref{fig:many_mode_aging}.
Uniform decay $\gamma=\kappa/2$ yields Eq.~\eqref{eq:spectral_delay_product}, including the vacuum factor $1+\kappa/(2\lambda_0)$.
The predicted delays per decade, $\ln(10)L_k$, agree with the computed median slopes [Fig.~\ref{fig:appE_dwell_laws}(b)] within $1.6\%$ and lie inside the slopes' pointwise 95\% bootstrap intervals.
For plateaus $\{20\},\ldots,\{16\}$, the predicted values are $(96.8,148.9,244.0,415.1,1010.6)\,U^{-1}$ and the computed values are $(96.4,146.7,240.4,411.2,1000.8)\,U^{-1}$, with a small systematic overestimate.
This comparison tests the logarithmic envelope. Phase-dependent projections can still modulate individual dwells as $\varepsilon_0$ varies.

The spectral ratio $L_{k+1}/L_k=(\lambda_k+\kappa/2)/\lambda_{k+1}$ describes sensitivities to $\varepsilon_0$, not absolute dwells.
Absolute dwells also contain corrections from the support changes, the exit amplitude, and the transfer factors. The support changes last $150$--$220\,U^{-1}$, comparable with the first two median dwells, $180$ and $399\,U^{-1}$.

\subsection{\texorpdfstring{Logarithmic event spacing}{Logarithmic event spacing}}
\label{app:log_event_spacing}
If accumulated decay is approximately proportional to elapsed time $t_k=t_{\mathrm{in}}(S_k)$, while local rates and transfer corrections vary slowly, Eq.~\eqref{eq:appD_inherited_dwell} reduces to $\tau_k\equiv\tau(S_k)\simeq a+bt_k$, with $b>0$.
Including the duration $\tau_k^{\rm sc}$ of the intervening support change, $t_{k+1}=t_k+\tau_k+\tau_k^{\rm sc}$, so $\tau_{k+1}\simeq(1+b)\tau_k+b\tau_k^{\rm sc}$.
When support changes are short compared with dwells,
\begin{equation}
\begin{aligned}
 \tau_{k+1}&\simeq q\tau_k,\qquad q=1+b,\\
 t_k+\frac{a}{b}&\simeq\left(t_0+\frac{a}{b}\right)q^k,\\
 k(t)&\simeq\frac{\ln[(t+a/b)/(t_0+a/b)]}{\ln q},\\
 \frac{dk}{dt}&\simeq\frac{1}{(t+a/b)\ln q}.
\end{aligned}
\label{eq:appD_log_event_spacing}
\end{equation}
Thus successive activations become approximately equally spaced in logarithmic time.
For uniform gain $\lambda$, $b=\kappa/(2\lambda)$, so $q=1+\kappa/(2\lambda)$ equals the spectral ratio of Appendix~\ref{app:spectral_delays}. For Fig.~\ref{fig:many_mode_aging}, fitting the five median dwells to $U\tau(S_k)=Aq^k$ gives $q=2.075$ with $R^2=0.995$ in log dwell [Fig.~\ref{fig:appE_dwell_laws}(c)]. This exceeds the decay-only estimate $1+\kappa/(2\lambda)\approx1.65$ for gains $\lambda\approx0.04U$: early support changes and the variation of local gains shift the dwell ratios [Fig.~\ref{fig:appE_dwell_laws}(d)]. The fitted $q$ characterizes this finite avalanche only.

\subsection{\texorpdfstring{Cyclic avalanches through the vacuum}{Cyclic avalanches through the vacuum}}
\label{app:recurrent_avalanches}
The loop of Fig.~\ref{fig:complex_cascades}(b) visits five saddles: the vacuum and the monomodes $\{20\},\{19\},\{18\},\{17\}$ on their $\sigma=-1$ branches.
Each has one unstable direction, carried by its escape mode.
Within the invariant subspace of the plateau mode and the escape mode, both branches of each unstable direction reach the next saddle, which is a sink in that subspace (Appendix~\ref{app:numerical_methods}).
These connections fix the five-plateau avalanche word at the chosen parameters, which the local spectra alone do not determine.

\begin{figure}[tb]
\centering
    \includegraphics[width=\columnwidth]{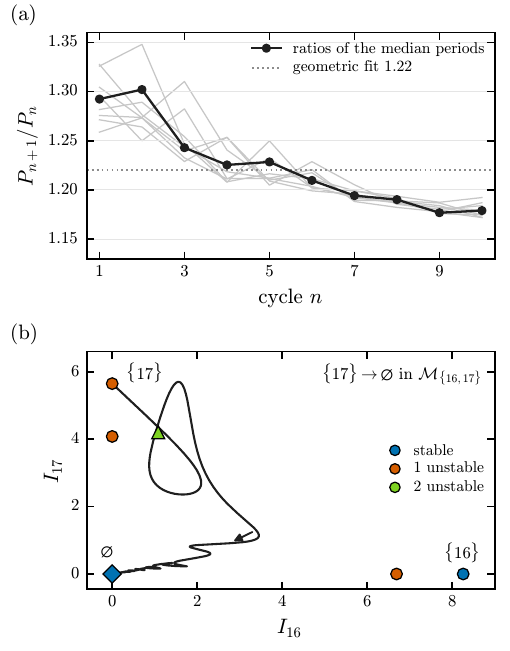}
\caption{Cyclic avalanche of Fig.~\ref{fig:complex_cascades}(b).
(a)~Successive period ratios for eight trajectories (gray) and ratios of the median periods (black), where $P_n$ is the duration of the $n$th loop; dotted line, the geometric fit $1.22$ over the observed window.
(b)~Connection $\{17\}\to\varnothing$ in $\mathcal M_{\{16,17\}}$, projected onto modal intensities.
FPs are colored by their number of unstable directions within this subspace (blue, zero; orange, one; green, two); symbols follow Fig.~\ref{fig:appC_global_organization}(b).
The connection passes the mixed saddle-focus at minimum relative phase-space distance $0.23$.
Parameters as in Fig.~\ref{fig:complex_cascades}(b).}
\label{fig:appE_recurrent_cycle}
\end{figure}

The loop periods grow with a ratio that varies across the observed window [Fig.~\ref{fig:appE_recurrent_cycle}(a)]. A geometric fit over this window gives $1.22$ per loop, consistent with the $7.5$-fold growth over eleven loops quoted in Sec.~\ref{sec:many_mode_aging}.

In particular, the connection from $\{17\}$ returns to the vacuum rather than the competing sink $\{16\}$ [Fig.~\ref{fig:appE_recurrent_cycle}(b)].
This connection reproduces the mixed-support episode of duration about $464\,U^{-1}$, but stays at relative phase-space distance at least $0.23$ from the mixed saddle-focus.
The episode on $\{16,17\}$ is therefore a nonlinear support change, not a dwell near that FP.

\FloatBarrier

\Needspace{10\baselineskip}
\section{\texorpdfstring{Numerical methods and coherent seed floor}{Numerical methods and coherent seed floor}}
\label{app:numerical_methods}
This appendix collects the numerical methods behind Figs.~\ref{fig:many_mode_aging}--\ref{fig:appE_recurrent_cycle} and quantifies the seed floor imposed by a coherent drive.

\subsection{\texorpdfstring{Numerical evolution and validation}{Numerical evolution and validation}}
\label{app:integration}

We integrate the deterministic mean-field equations with fixed-step fourth-order Runge--Kutta (RK4) in double precision.
Log-intensity/phase coordinates retain the strongly decayed dormant amplitudes in the many-mode maps and aging trajectories; Cartesian quadratures are used for Fig.~\ref{fig:complex_cascades}(a).
Initial directions are uniform on the real $2N$-dimensional sphere, obtained by normalizing independent Gaussian vectors and multiplying by $\varepsilon_0$.
Table~\ref{tab:appE_integration} collects the integration settings.
The maps use 32 directions per point. The aging ensemble has 128 directions, and the cyclic avalanche has eight.
For Fig.~\ref{fig:many_mode_aging}(c)--(f), $(J,\Delta,\chi,\kappa)/U=(1,1.9885,0.067,0.05695)$; for Fig.~\ref{fig:complex_cascades}(b), these values are $(1,2,0.056,0.028)$, with $G=\sqrt{\chi^2+(\kappa/2)^2}$ in both cases.

\begin{table}[htb]
\caption{Integration settings. ``Log'' denotes log-intensity/phase coordinates and ``Cart.'' real Cartesian quadratures.}
\label{tab:appE_integration}
\centering
{\small
\begin{tabular}{lccc}
\hline\hline
Calculation & Coordinates & $U\,dt$ & $UT$ \\
\hline
Maps, Fig.~\ref{fig:many_mode_landscapes} & Log & $0.02$ & $5000$ \\
Aging, Fig.~\ref{fig:many_mode_aging} & Log & $0.02$ & $5\times10^4$ \\
Fig.~\ref{fig:complex_cascades}(a) & Cart. & $0.05$ & $10^5$ \\
Fig.~\ref{fig:complex_cascades}(b) & Log & $0.02$ & $2\times10^5$ \\
\hline\hline
\end{tabular}}
\end{table}

With $U\,dt=0.01$, all 128 deterministic aging trajectories retain their avalanche word and plateau classification; the largest change in a transient-dwell median is $1.93\times10^{-3}U^{-1}$.
A deterministic Cartesian-coordinate control also retains the word and terminal mode for all directions, with relative median-dwell changes below $8.9\times10^{-4}$.

Doubling the horizons of Fig.~\ref{fig:complex_cascades} preserves the four-mode support and its two spectral peaks in panel~(a), and the cyclic word in all eight trajectories of panel~(b).
For the spectra, we subtract the mean and taper the intensity signal at both ends of four consecutive windows of length $5000U^{-1}$.
The same dominant frequency bin appears for every active mode in every window, separately in $0.002\le f/U<0.05$ and $0.10\le f/U<0.25$.
The resulting peaks, $f/U=0.0116$ and $0.1766$, are resolved to $2\times10^{-4}U$.

\subsection{\texorpdfstring{Periodic orbits and saddle connections}{Periodic orbits and saddle connections}}
\label{app:periodic_orbits}
Periodic branches in Fig.~\ref{fig:appC_global_organization}(a) use multiple shooting with pseudo-arclength continuation.
DOP853 integrates the state and variational equations with relative and absolute tolerances $3\times10^{-10}$ and $10^{-12}$, and maximum step $U\,dt=0.05$.
Shooting closure is at most $10^{-8}$. For Floquet stability, we exclude the neutral multiplier nearest unity.
For the five connections of Appendix~\ref{app:deterministic_methods}, we launch both branches of each saddle's unit unstable eigenvector at displacements $10^{-6}$ and $10^{-8}$, retaining only the plateau mode and the escape mode.
All twenty launches reach the next saddle as a sink in the retained invariant subspace.
Distances to the mixed FP are evaluated in real quadratures and minimized over its parity images.

\subsection{\texorpdfstring{Outcome classification, plateau detection, and exit amplitudes}{Outcome classification, plateau detection, and exit amplitudes}}
\label{app:plateau_detection}

The maps in Fig.~\ref{fig:many_mode_landscapes} classify finite-time outcomes.
At the final time $T$, with state $\boldsymbol r_T=\boldsymbol r(T)$, $I_m=u_m^2+v_m^2$, $I_{\max}=\max_m I_m$, and
\begin{equation}
 I_{\rm cut}=\max(10^{-6},10^{-2}I_{\max}),
 \qquad S_T=\{m:I_m>I_{\rm cut}\}.
\end{equation}
We classify the endpoint as vacuum if $\lVert\boldsymbol r_T\rVert_2\le2\times10^{-5}$ or $S_T=\varnothing$.
The saved tail comprises 4097 samples over the last $10\%$ of the trajectory.
For this tail, let $I_{\rm sc}=\max(\max_{\rm tail}\sum_m I_m,10^{-12})$ and define
\begin{equation}
\begin{aligned}
 W_m^{(h)}&=\frac{\mathrm{pct}_{95}^{(h)}(I_m)-\mathrm{pct}_{5}^{(h)}(I_m)}{I_{\rm sc}},\\
 \varrho_m&=\frac{W_m^{({\rm late})}}
 {\max[W_m^{({\rm early})},10^{-2},10^{-14}/I_{\rm sc}]},\\
 d_{\rm tail}&=\frac{\lVert\langle\boldsymbol I\rangle_{\rm last\ fifth}
 -\langle\boldsymbol I\rangle_{\rm first\ fifth}\rVert_2}{I_{\rm sc}},
\end{aligned}
\end{equation}
where $\mathrm{pct}_p^{(h)}$ is the $p$th intensity percentile on interval $h$; early and late refer to the two tail halves.
Here, $\dot{\boldsymbol r}=\boldsymbol F(\boldsymbol r)$ denotes Eq.~\eqref{eq:mf_general_complex} in real quadratures.
Table~\ref{tab:appE_outcomes} lists the criteria for the stationary, persistent, and unresolved labels.
\begin{table}[htb]
\caption{Outcome labels of the maps in Fig.~\ref{fig:many_mode_landscapes} and their criteria at the final time $T$; rates in units of $U$.}
\label{tab:appE_outcomes}
\centering
{\small
\begin{tabular}{ll}
\hline\hline
Label & Criteria \\
\hline
Stationary & \parbox[t]{0.62\columnwidth}{$\lVert\boldsymbol F(\boldsymbol r_T)\rVert_2/\max(1,\lVert\boldsymbol r_T\rVert_2)\le10^{-2}$; $W_m^{(\mathrm{full})}\le0.05$ for the mode selected by the tail diagnostic; $d_{\rm tail}\le0.05$; $\max\operatorname{Re}\operatorname{spec}\mathcal J_{\rm R}(\boldsymbol r_T)\le10^{-3}$} \\
Persistent, short period & \parbox[t]{0.62\columnwidth}{one mode with $W_m^{(\mathrm{late})}\ge10^{-2}$, $\ge10$ full-tail and $\ge5$ late-half extrema, and $\varrho_m\ge0.999$} \\
Persistent, long period & \parbox[t]{0.62\columnwidth}{one mode with $\ge32$ tail samples, $W_m^{(\mathrm{early})},W_m^{(\mathrm{late})}\ge10^{-2}$, $\ge4$ full-tail and $\ge2$ late-half extrema, and $\varrho_m\ge0.25$} \\
Unresolved & \parbox[t]{0.62\columnwidth}{none of the above; the endpoint support $S_T$ is kept} \\
\hline\hline
\end{tabular}}
\end{table}

FP candidates from Eq.~\eqref{eq:appB_support_branch} are refined to residual at most $10^{-7}$ and classified with the full real Jacobian; mixed supports are not exhaustively enumerated in large networks.

For avalanche trajectories, we define the effective support $S_{\rm eff}(t)$ as the smallest set carrying $99.75\%$ of the total intensity.
A certified plateau retains this support within relative distance $d_{{\rm FP},S}=\min\lVert\boldsymbol r-\boldsymbol r_S\rVert_2/\lVert\boldsymbol r_S\rVert_2\le0.05$, minimized over the FP's parity images.
Confirmation requires remaining inside for at least $5U^{-1}$, at least three saved samples, and total duration $U\tau\ge20$.
Entry and exit use the same distance threshold and interpolated crossings.
This criterion certifies sustained proximity, not an invariant connection.
Nonterminal plateaus surviving to $T$ are right-censored at $T-t_{\rm in}$; stable terminal plateaus are excluded from escape statistics.

Local escape tests use the eigenvector normalization of Appendix~\ref{app:dimer}: a unit right eigenvector in real quadratures and its dual left row.
For $N=20$, the incoming seed is projected at the first saved sample inside the certified plateau.
The dwell $\tau$ plotted in Fig.~\ref{fig:many_mode_aging}(d) is the certified plateau duration, bounded by the interpolated distance crossings above.
The exit amplitudes are $c_\star=0.1\sqrt{I_o}$ for the dimer and $0.05\sqrt{I_m}$ for plateau $\{m\}$ of the $N=20$ avalanche.
The $N=20$ exit amplitude matches the relative distance criterion $0.05$, although the projected coordinate and the full distance need not cross these values at the same time.
\subsection{\texorpdfstring{Ensemble statistics and persistence}{Ensemble statistics and persistence}}
\label{app:persistence_statistics}

The representative trajectory in Fig.~\ref{fig:many_mode_aging}(c) minimizes the squared distance from the five median log-dwells on plateaus $\{20\}$--$\{16\}$. These medians are $(180,399,769,1426,3665)\,U^{-1}$.
The entry-aligned persistence curves use one resolved dwell per trajectory and plateau: at elapsed time $t$, persistence is the fraction exceeding $t$.
Waiting-time observations require a resolved outcome on a nonterminal plateau and at least 16 distinct trajectories in each $(t_w,t/t_w)$ cell, also conditioned on the plateau when appropriate.
We exclude support changes, terminal plateaus, and censored outcomes from the fits.

The mixture over successive plateaus is fitted by Eq.~\eqref{eq:global_kww_fit}, giving $\mu_P=0.98$ with 95\% interval $[0.97,1.00]$.
The plateau-conditioned fit instead uses
\begin{equation}
 \widehat P_m(t_w,t)=\exp\!\left[-\left(
 \frac{t}{z_{0,m}t_w^{\mu_P^{\rm plat}}}\right)^{\beta_m}\right],
\end{equation}
with a shared exponent but separate $z_{0,m}$ and $\beta_m$ for each plateau.
It gives $\mu_P^{\rm plat}=-1.440$, with 95\% interval $[-1.444,-1.437]$.
The negative exponent is consistent with decreasing remaining lifetimes within plateaus [Fig.~\ref{fig:many_mode_aging}(f)], whereas the positive pooled exponent reflects progression to slower plateaus.
Intervals use 500 bootstrap resamples of complete trajectories, retaining their correlated observations.
Eightfold cross-validation keeps all observations of a trajectory together, trains on seven folds, and compares held-out mean Bernoulli negative log likelihood with the restriction $\mu_P=0$.
The pooled fit improves the held-out likelihood over $\mu_P=0$ for all 128 trajectories, and the plateau-conditioned fit for 118 of them (92\%).

The global exponent does not imply a single curve $P(t_w,t)=\mathcal F(t/t_w)$.
For example, at the same relative lag $t/t_w=1/4$, persistence is $0.94$ at $Ut_w\simeq2263$ and $0.10$ at $Ut_w\simeq3200$ (115 and 126 at-risk trajectories, respectively).
The sawtooth associated with passage through successive plateaus remains after rescaling.

Dwell and local-escape summaries use 5000 whole-trajectory bootstrap samples.
Spectral-delay slopes use 2000, preserving the pairing of directions across the five initial amplitudes $\varepsilon_0=10^{-2},\ldots,10^{-6}$.

\subsection{\texorpdfstring{Coherent seed floor}{Coherent seed floor}}
\label{app:coherent_seed}
A uniform coherent one-photon drive of real amplitude $\eta$ adds $\eta$ to the right-hand side of Eq.~\eqref{eq:mf_general_complex} for every mode.
When this source feeds the outgoing channel, it produces a seed proportional to $\eta$ at leading order, hence slope $-1$ in $\lambda_+\tau$ versus $\ln\eta$.
For the $N=20$ avalanche at $\eta/U\simeq5\times10^{-8}$, all 32 sampled trajectories retain the six-plateau word, while the ratio of the median dwells on $\{16\}$ and $\{20\}$ falls from $20.3$ to $2.0$.
A static local-response estimate captures the slope but not the absolute dwells, which also depend on forcing accumulated before entry.
The drive thus compresses the dwell growth before it changes the avalanche word.

\end{document}